\documentclass[useAMS,usenatbib]{mn2e}

\usepackage{rotating}
\usepackage{longtable}
\usepackage{graphics}
\usepackage{aas_macros}

\newcommand{\teff}{$T_{\rm eff}$}
\newcommand{\eexc}{$E_{\rm exc}$}

\def\vt{$\xi_{\rm t}$}

\newcommand{\kms}{km\,s$^{-1}$}

\def\ione{\,{\sc i}}
\def\ii{\,{\sc ii}}
\def\iii{\,{\sc iii}}
\newcommand{\kH}{$S_{\rm H}$}    
\newcommand{\dnlte}{$\rm \Delta_{NLTE}$}

\title[A NLTE line formation for Ca\ione\ and Ca\ii\ in model atmospheres of  B-F stars]{A NLTE line formation for neutral and singly-ionised calcium in model atmospheres of B-F stars}
\author[T. M. Sitnova, L. I. Mashonkina, T. A. Ryabchikova]{T. M. Sitnova$^1$\thanks{E-mail:
sitnova@inasan.ru}; L. I. Mashonkina$^{1, 2}$, T. A. Ryabchikova$^1$ \\
1-Institute of Astronomy, Russian Academy of Sciences, Pyatnitskaya 48, 119017, Moscow, Russia\\
2-Department of Theoretical Physics, A. I. Herzen University, St. Petersburg 191186, Russia\\
}
\begin{document}

\date{}

\pagerange{\pageref{firstpage}--\pageref{lastpage}} \pubyear{2018}

\maketitle

\label{firstpage}

\begin{abstract}
We present non-local thermodynamic equilibrium (NLTE) line formation calculations for  Ca\ione\ and Ca\ii\ in B-F stars. 
The sign and the magnitude of NLTE abundance corrections depend on line and stellar parameters. 
We determine calcium abundances for nine stars with reliable stellar parameters. 
For all stars, where the lines of both species could be measured, the NLTE abundances are found to be consistent within the error bars.
We obtain consistent NLTE abundances from Ca\ii\ lines in the visible and near infra-red (IR, 8912-27, 9890 \AA) spectrum range, in contrast with  LTE, where the discrepancy between the two groups of lines  ranges from $-0.5$~dex to 0.6~dex for different stars. 
Our NLTE method reproduces the  Ca\ii\ 8912-27, 9890 \AA\ lines observed in  emission in the late B-type star HD~160762  with the classical plane-parallel and LTE model atmosphere.
NLTE abundance corrections  for lines of Ca\ione\ and Ca\ii\   were calculated in a grid of model  atmospheres with 7000~K $\leq$ \teff\ $\leq$ 13000~K, 3.2 $\leq$ log~g $\leq$ 5.0, $-0.5 \leq$ [Fe/H] $\leq 0.5$, \vt = 2.0 \kms. 
Our NLTE results can be applied for calcium NLTE abundance determination from Gaia spectra, given that accurate continuum normalisation and proper treatment of the hydrogen Paschen lines are provided. 
The NLTE method can be useful to refine calcium underabundances in Am stars and to provide accurate observational constraints on the models of diffusion.

\end{abstract}


\section{Introduction}

Calcium is  observed in the two ionisation stages, Ca\ione\ and Ca\ii, in wide range of stellar parameters. This provides an opportunity to determine atmospheric parameters and also calcium abundance using Ca\ione\ and Ca\ii\ lines. 
Therefore, one needs to explore how deviations from LTE influence on formation of Ca\ione\ and Ca\ii\ lines.

Statistical equilibrium of Ca\ione\--\ii\ in late type stars is explicitly  studied in the literature,  for example  by \citet{Drake1991,Thevenin2000}, and \citet[][and references therein]{mash_ca,mash_ca_2017}.

In the  early spectral type atmospheres  with effective temperatures of 15000~K $\le$ \teff\ $\le$ 20000~K and surface gravity of log~g from 2.5 to 4.0, the NLTE line formation for Ca\ii\ 3933 \AA\ line was investigated by \citet{Mihalas1973}. 
In NLTE, he found the  line strengthening, by a factor of  about two and five, in terms of abundances, for main sequence stars and giants, respectively. 

In the  \teff\ range from 8000~K to 13000~K, the NLTE calculations for  Ca\ione\--\ii\ were not yet performed, although observations reveal signatures of strong deviations from LTE.
 \citet{2004A&A...418.1073W} mentioned the presence of the emission Ca\ii\ lines of stellar origin in the spectrum of He-weak star 3~Cen~A.
In general, the emission lines can not be explained within a classical  LTE approach.
\cite{Alexeeva_c_hot}, for example, evaluated the NLTE line formation for C\ione-\ii\ in atmospheres of B-F stars and explained a mechanism of C\ione\ emission  line formation using plane-parallel and LTE model atmosphere.

The NLTE calculations for calcium are expected to be important for chemically peculiar stars, in particular, for Am (melallic-line) stars, which show  underabundances of calcium and scandium and an overabundance of iron peak and  rare-earth elements \citep{1974ARA&A..12..257P}.
Am stars are located on the main sequence with log~g from 4.5 to 3.5. \citet{1992A&A...253..451B} examined behaviour of calcium abundance  in relation to the evolutionary stage of some Am stars in three star clusters and found that the more the Am star is evolved, the less is its calcium deficiency. In contrast to previous studies, from analysis of calcium abundance in field Am stars, \citet{1998A&A...330..651K} found that calcium abundance does not tend to be larger in evolved Am stars than in unevolved ones, for objects distributed along a given evolutionary track in the HR diagram. However, they  found a significant correlation between calcium abundance and effective temperature, in the sense that the cooler objects are the most Ca-deficient, hence have the most pronounced Am peculiarity. The latter result was confirmed by \citet{2005AstL...31..388R}. The above mentioned studies rely on assumption of local thermodynamic equilibrium (LTE).

In this study, we investigate the NLTE effects for Ca\ione\--\ii\ in atmospheres of BAF stars.
This paper aims to understand a mechanism of the Ca\ii\ emission in B-type stars and to treat the method of accurate abundance determination for BAF stars based on the NLTE line formation for Ca\ione\--\ii.

We describe the method of calculations in Section~\ref{method}. 
Section~\ref{se} shows NLTE effects for  Ca\ione\ and Ca\ii\ depending on atmospheric parameters.
Our stellar sample and their NLTE and LTE calcium abundances are presented in Section~\ref{hots}.
Our recommendations and conclusions are given in Sect.~\ref{con}. 

\section{Method of NLTE calculations for Ca\ione\--\ii}
\label{method}


We rely on  a fairly complete model atom constructed by \citet{mash_ca}. The model atom
 includes 63 levels of Ca\ione, 37 levels of Ca\ii, and the ground state of Ca\iii. 
 Energy levels up to 0.17/0.67~eV below the ionization threshold are included in the Ca\ione/Ca\ii\ model atom.
 For Ca\iii, only the ground state is considered because the first excited level has \eexc = 24~eV and its population is tiny compared to that of the ground state.
Fine structure splitting sub-levels are included explicitly for the 
Ca\ione\ 4p$^3$P$^\circ$ and 3d$^3$D
and Ca\ii\ 3d$^2$D, 4p$^2$P$^\circ$ and 4d$^2$D terms.
For radiative transitions, accurate data on
photoionisation cross-sections and transition probabilities from the Opacity
Project \citep[OP; see][for a general review]{1994MNRAS.266..805S} were used, which are accessible in the TOPBASE\footnote{http://cdsweb.u-strasbg.fr/topbase/topbase.html} database. 
For electron impact excitation, detailed results from the $R-$matrix calculations are available for ten transitions from the ground state in Ca\ione\ \citep{2001ADNDT..77...87S} and all the transitions between levels of the $n \le 8$ configurations of Ca\ii\ \citep{ca2_bautista}. 
For the remaining bound-bound transitions, approximate formulae are used, namely, the formula of \citet{Reg1962} for the allowed transitions and the formula from \citet{WA1948_cbb} with a collision strength of 1.0 for the radiatively forbidden transitions.
Ionisation by electronic collisions is calculated with the \cite{1962amp..conf..375S} formula using the threshold photoionisation cross-section from hydrogenic approximation.

In the model atmospheres with \teff\ $\le$ 7500~K, the statistical equilibrium (SE) calculations take into account inelastic collisions with neutral hydrogen atoms using the classical \citet{Drawin1968,Drawin1969} approximation as implemented by \citet{Steenbock1984}. The collision rates are scaled by a factor of \kH\ = 0.1 as recommended by \citet{mash_ca}.  
It is worth noting that, in the atmospheric parameter range with which we concern in this study, using approximate rates for Ca + H\ione\ collisions produces minor effect on the final NLTE results. 
For example, in the model with \teff / log~g = 7000/ 4.4, the abundance difference between neglecting and  including this type of collisions does not exceed 0.02~dex for lines of Ca\ione\ and Ca\ii\ in the visible spectrum range, however it amounts to 0.08~dex for Ca\ii\ 9890 \AA\ line, which is the most sensitive to the details of NLTE calculations. 

 The NLTE and synthetic spectrum calculations are performed with the codes described by \citet{Sitnova_ti}. 
As in \citet{Sitnova_ti}, model atmospheres were calculated with the code \textsc {LLmodels} \citep{llmod}. 
The exception is Sirius, for which model atmosphere was taken from R.~Kurucz website\footnote{http://kurucz.harvard.edu/stars/sirius/ap04t9850g43k0he05y.dat}. 

\section{Departures from LTE for Ca\ione-\ii}
\label{se}


\subsection{Statistical equilibrium of Ca\ione-\ii}
\label{se}

\begin{figure*}
	\centering
	\includegraphics[width=80mm]{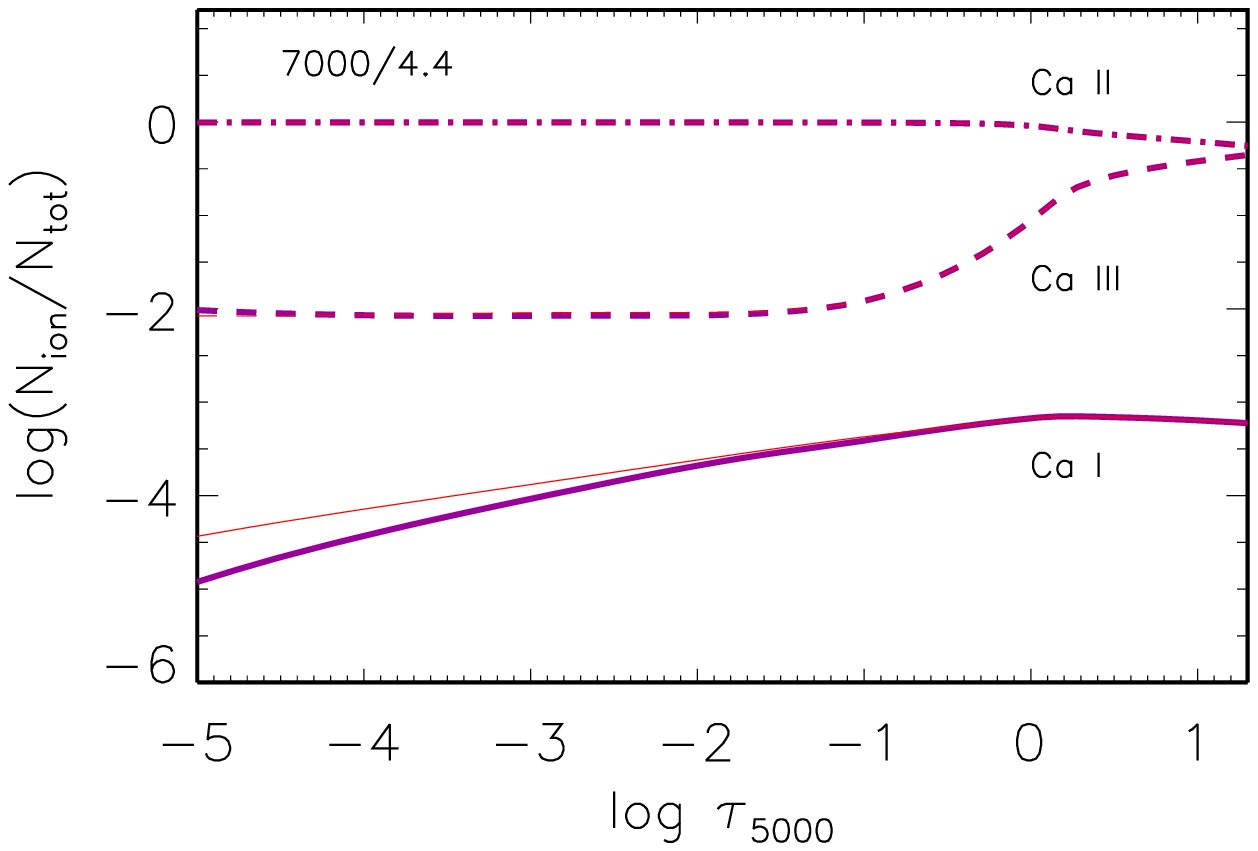}
	\includegraphics[width=80mm]{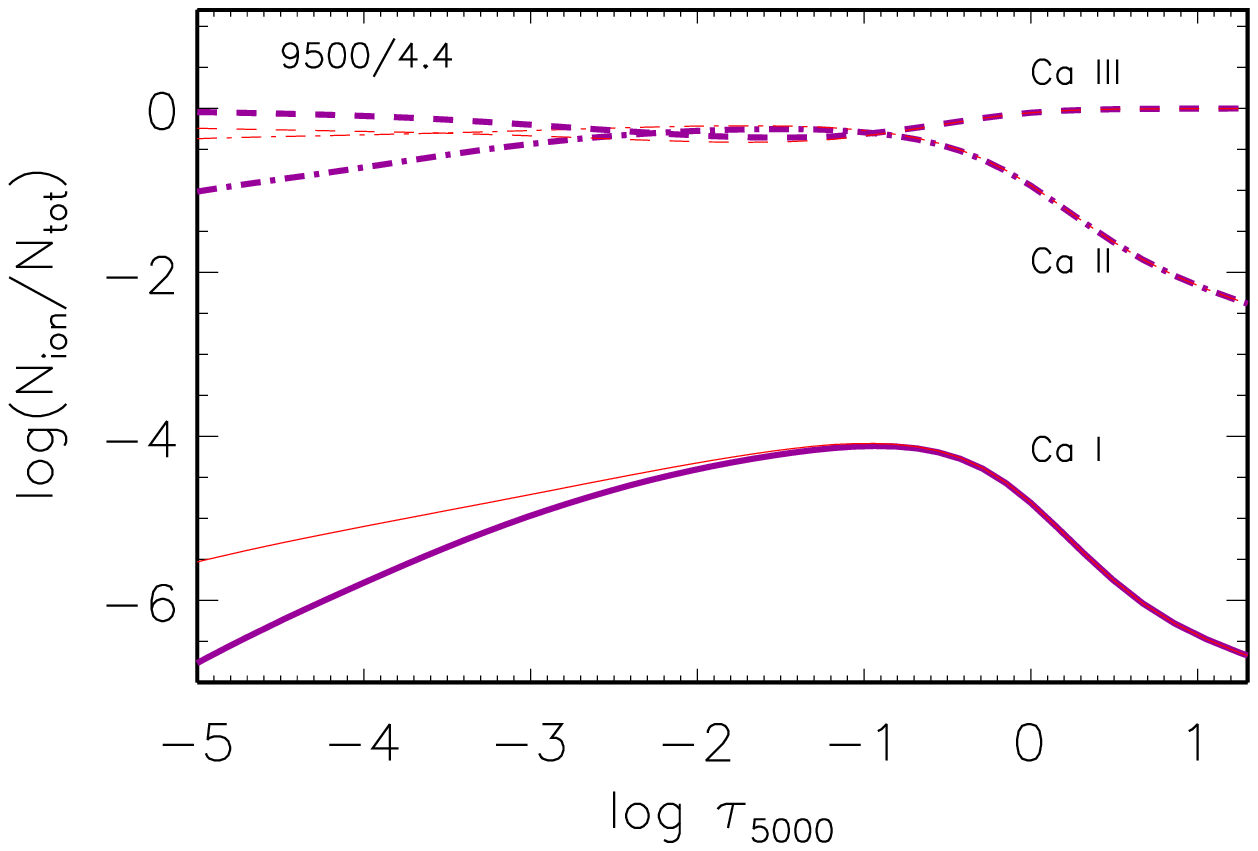}
	\includegraphics[width=80mm]{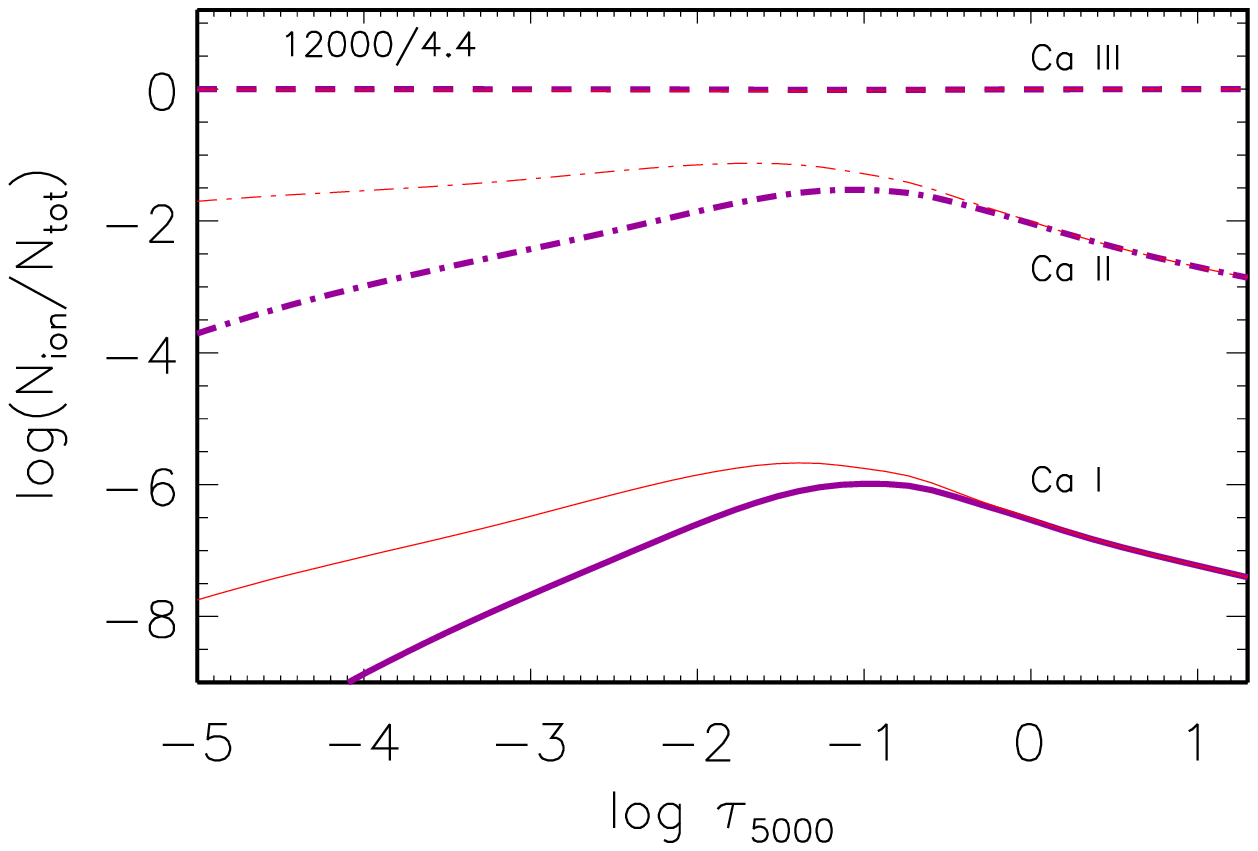}
	\includegraphics[width=80mm]{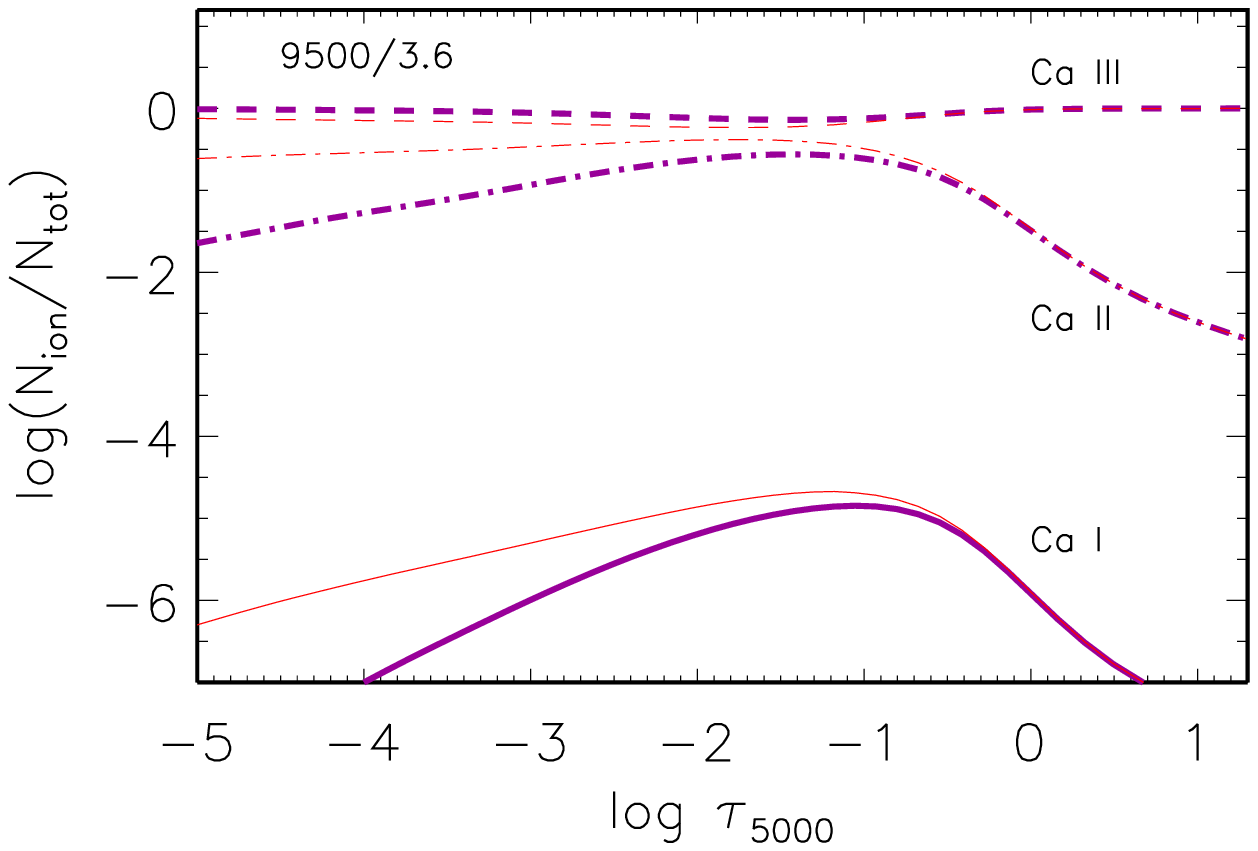}
	\caption{	
NLTE (thick lines) and LTE (thin lines) number density fractions of Ca\ione, Ca\ii, and Ca\iii\ (N$_{\rm ion}$/N$_{\rm tot}$)  in different model atmospheres. 
For each model, atmospheric parameters \teff / log~g are indicated.  We assume solar chemical composition everywhere.}		
	\label{ionisation} 
\end{figure*}

The NLTE and LTE number density fractions of Ca\ione, Ca\ii, and Ca\iii\  in different model atmospheres are presented in Fig.~\ref{ionisation} as a function of the continuum optical depth log~$\tau_{5000}$ (referring to $\lambda$ = 5000~\AA). 
In stellar atmospheres with \teff\ $\ge$ 7000~K, calcium is strongly ionised, with N(Ca\ione)/N(Ca\ii) $\leq$ 10$^{-4}$. 
NLTE results in underpopulation of Ca\ione\ relative to the thermodynamic equilibrium population in atmospheric layers above log~$\tau_{5000}$~=~0. In the deeper atmospheric layers, each species keeps its LTE fraction. 
In atmospheres with 7000~K $\le$ \teff\ $\le$ 8000~K, Ca\ii\ is a majority species, with N(Ca\ii)/N(Ca\iii) $\ge$ 10, which results in small deviations from
LTE in the total number density  of Ca\ii.
The ionisation degree increases toward higher \teff\ and lower log~g. 
In atmospheres with 9000~K $\le$ \teff\ $\le$ 9500~K, Ca\ii\ and Ca\iii\ are competing species, and favorable conditions for overionisation of Ca\ii\ appear.
With a further increase of \teff, Ca\iii\ becomes the majority species, and overionisation of Ca\ii\ increases.

\begin{figure*}
	\centering
	\includegraphics[width=80mm]{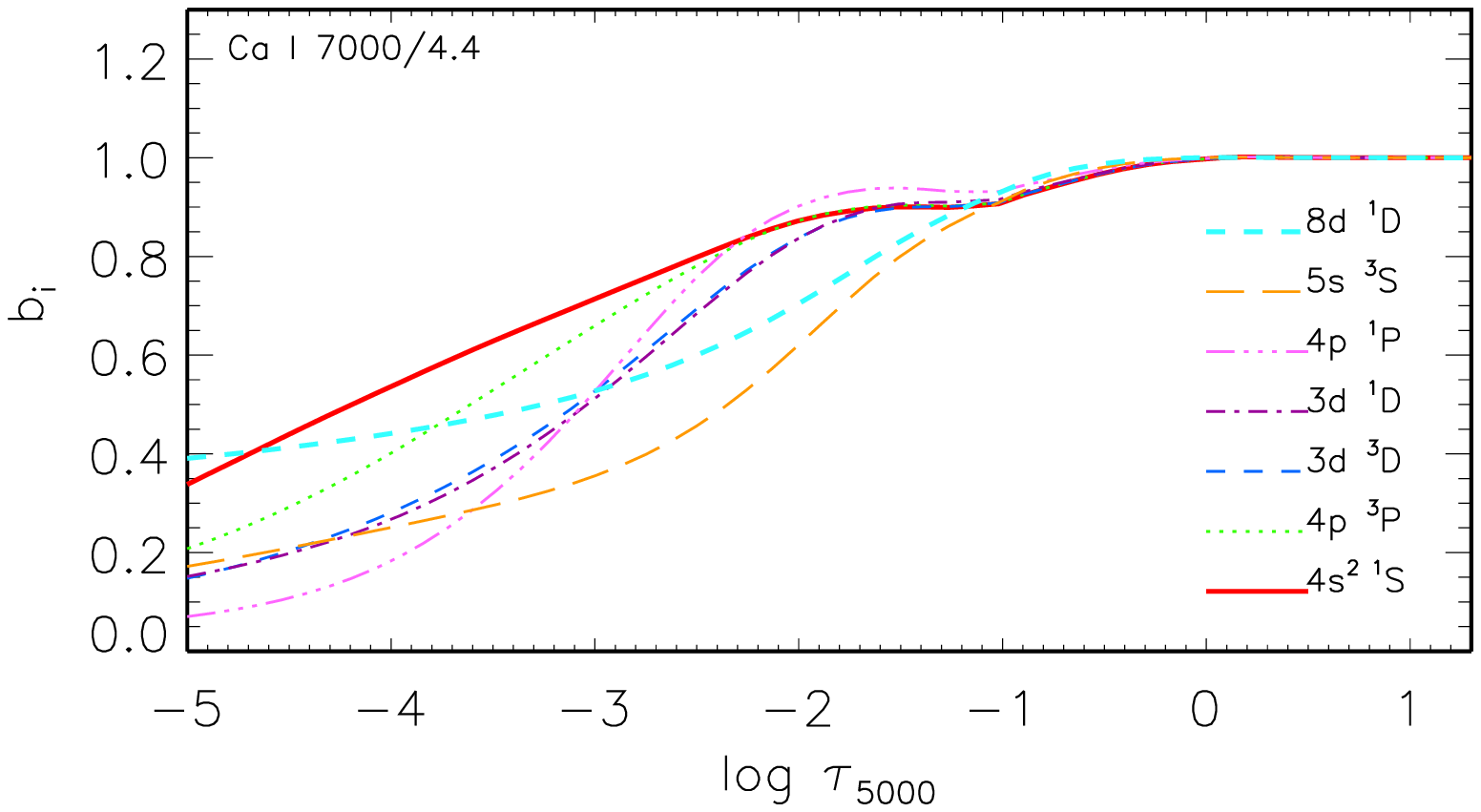}
	\includegraphics[width=80mm]{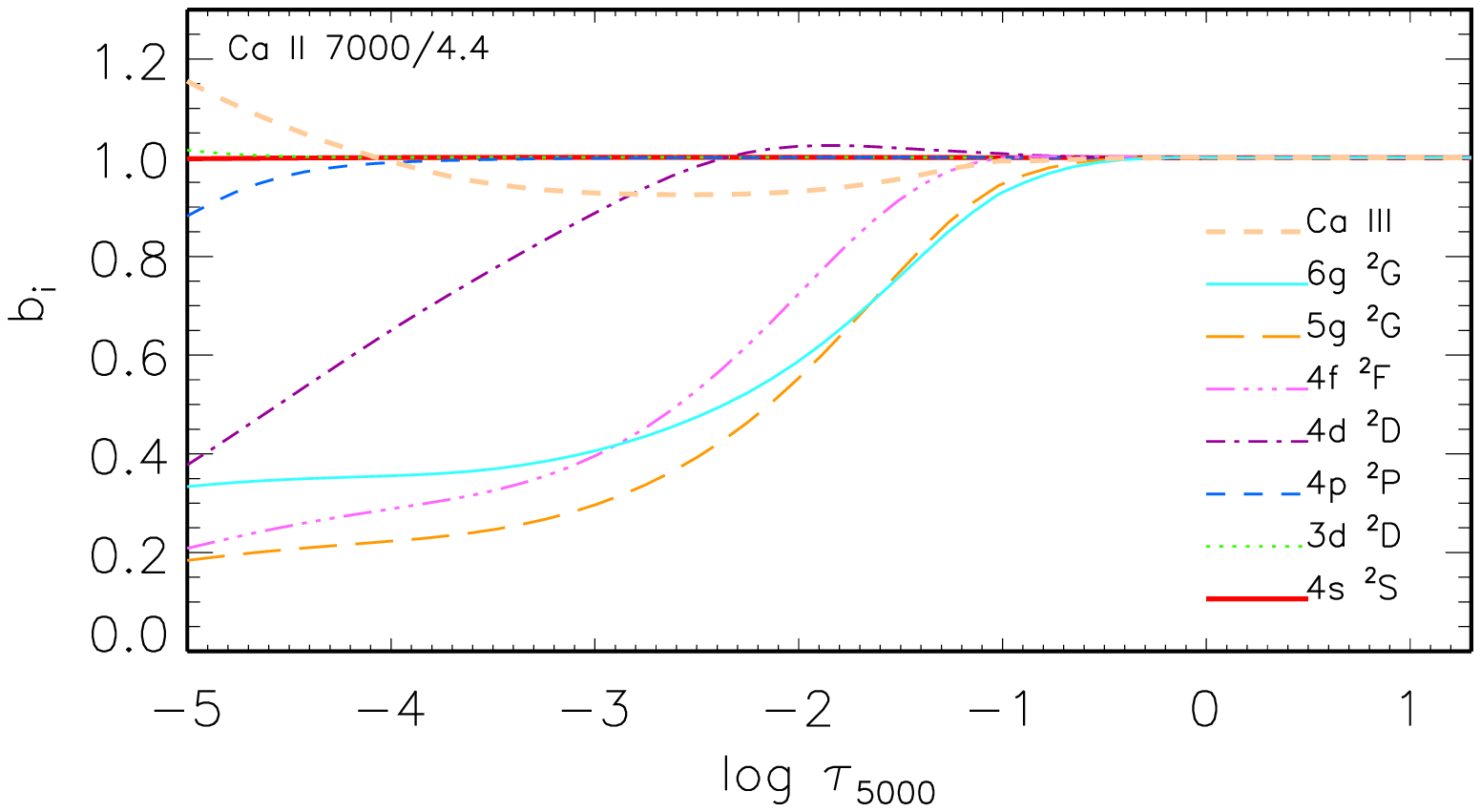}
	\includegraphics[width=80mm]{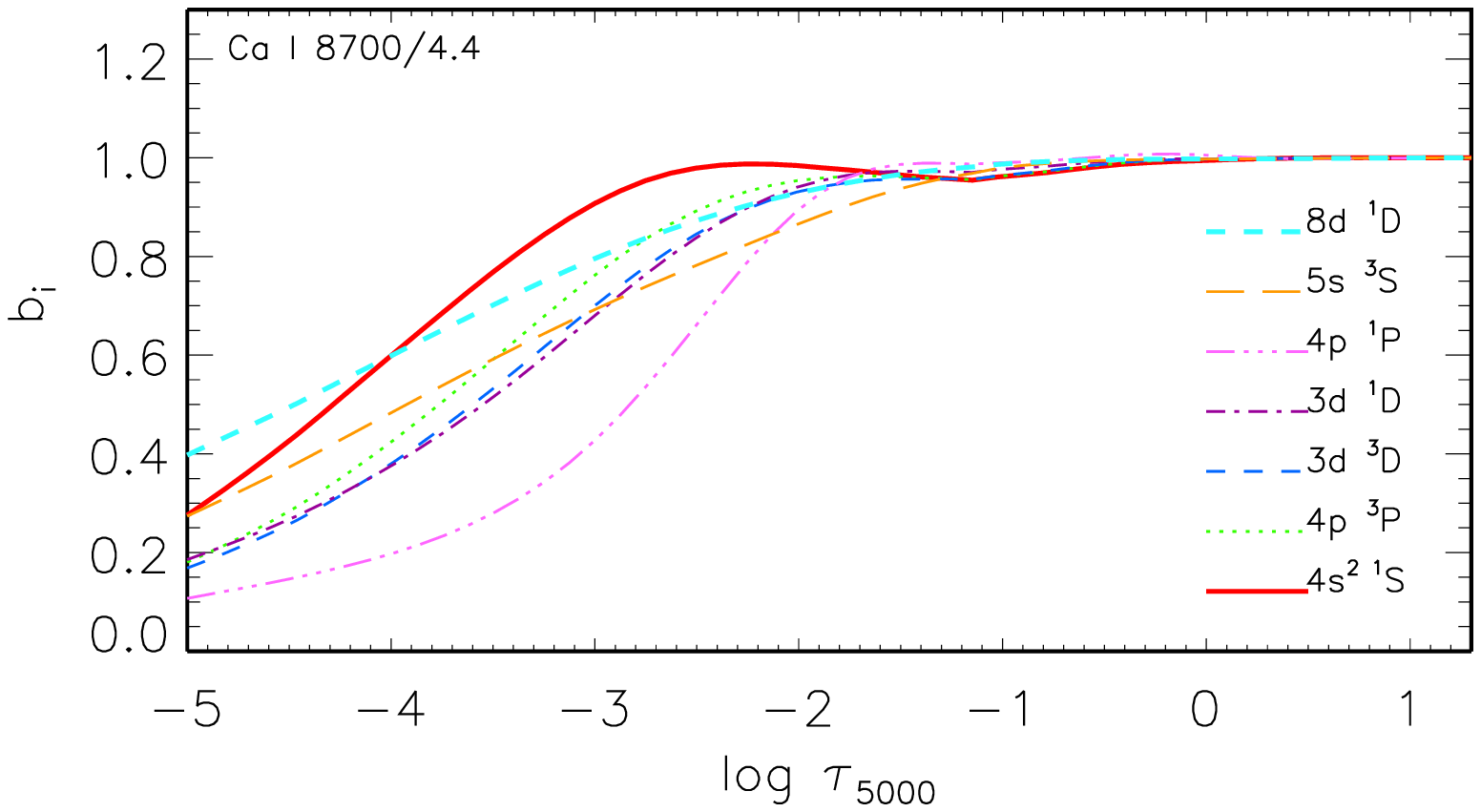}
	\includegraphics[width=80mm]{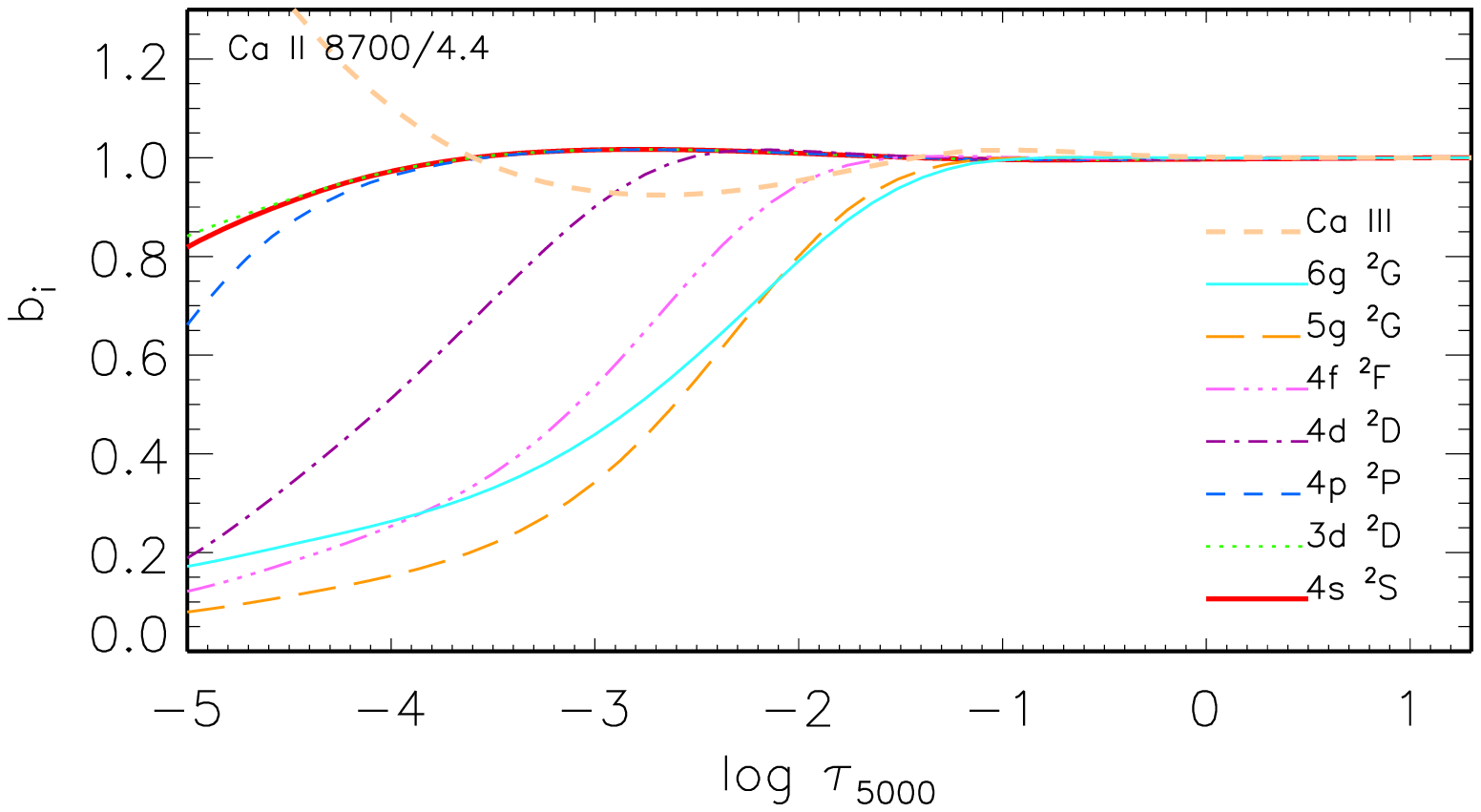}	
	\includegraphics[width=80mm]{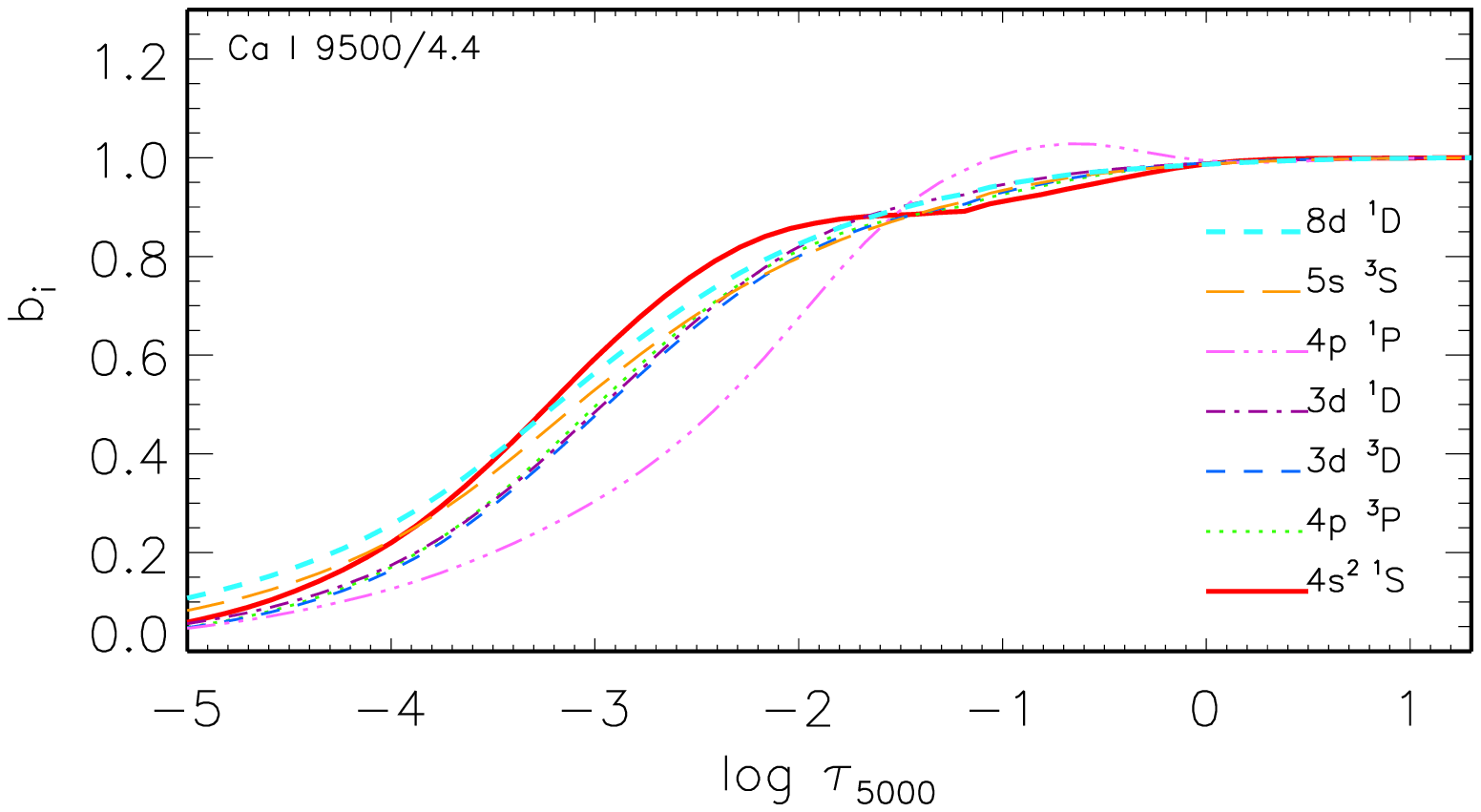}
	\includegraphics[width=80mm]{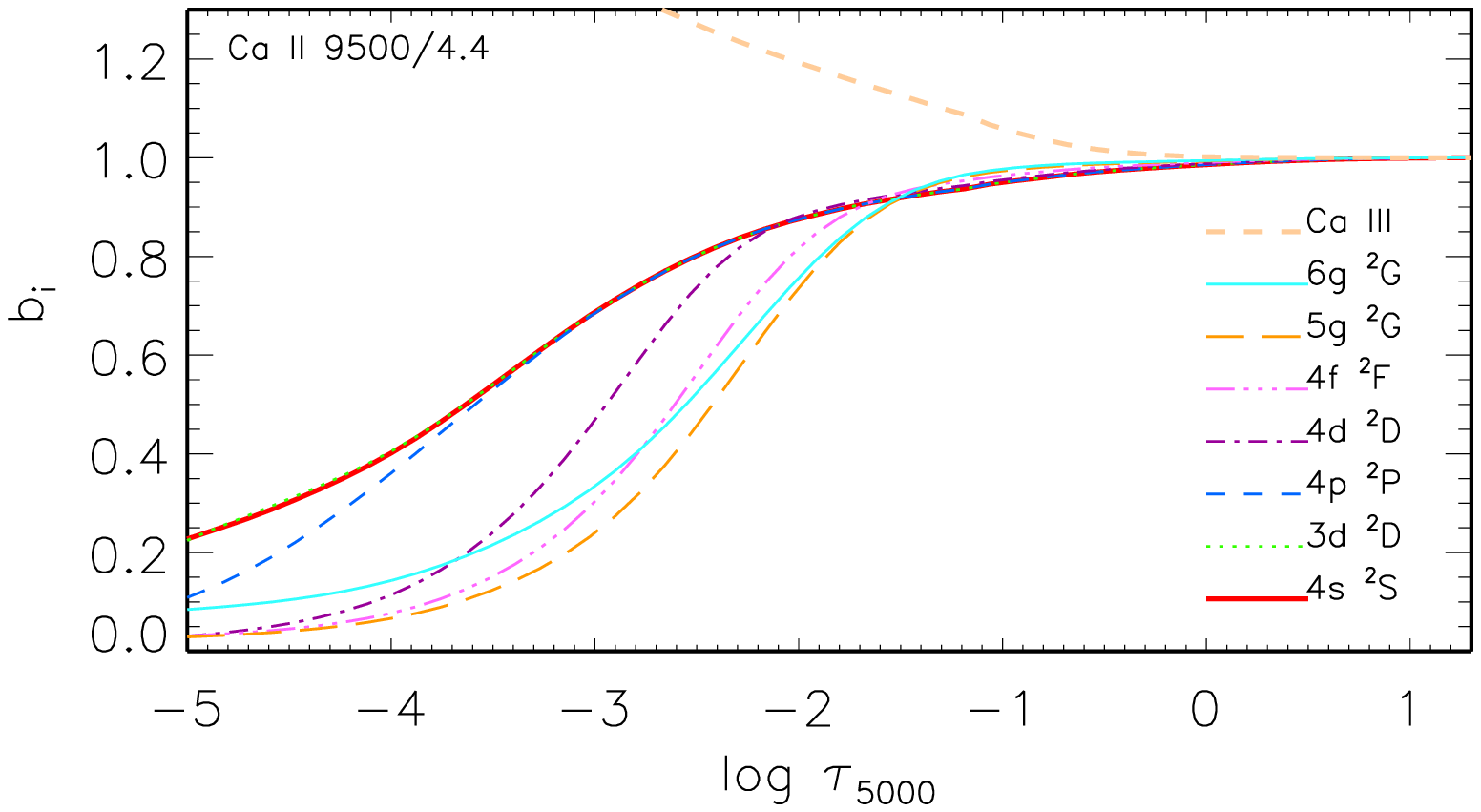}	
	\includegraphics[width=80mm]{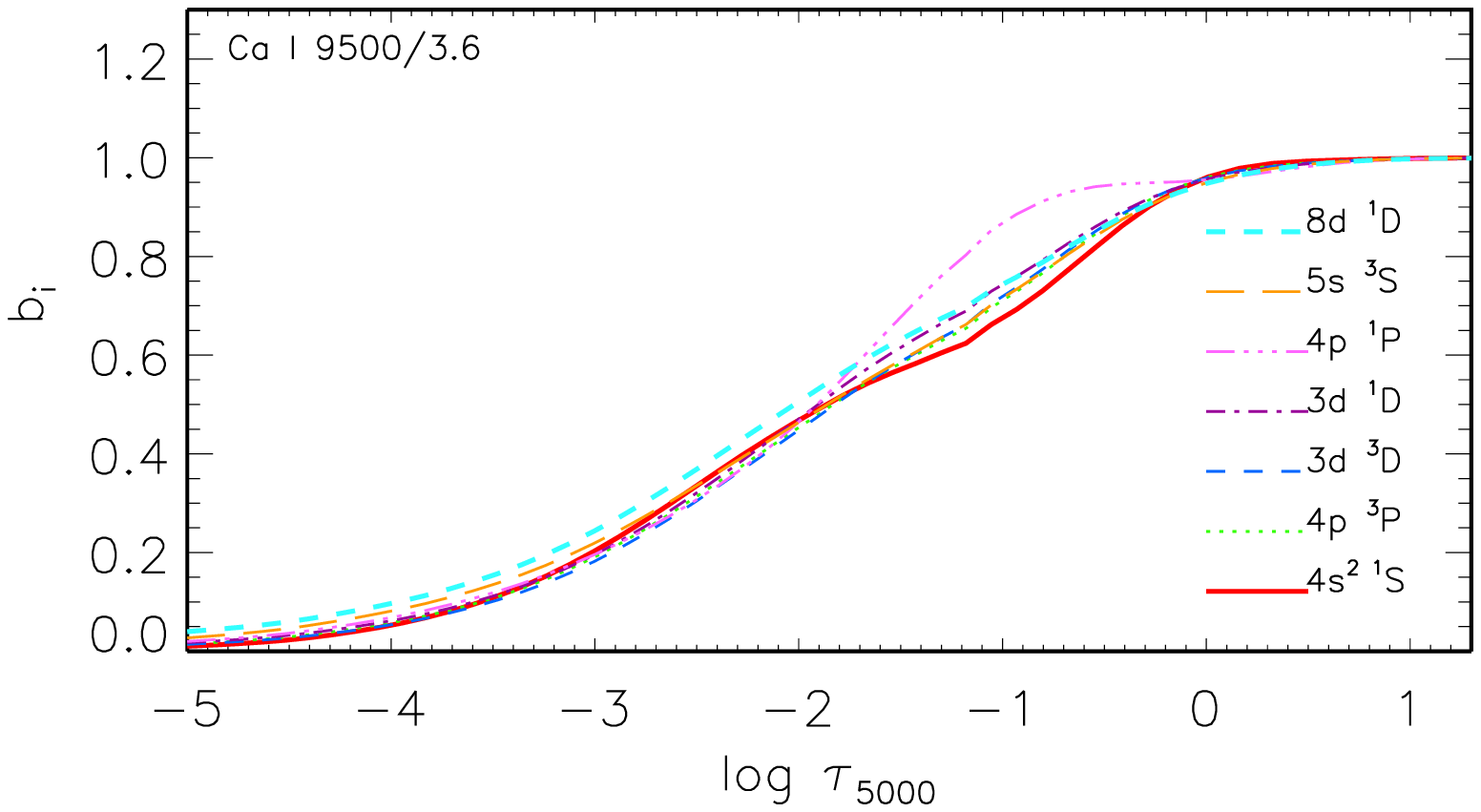}
	\includegraphics[width=80mm]{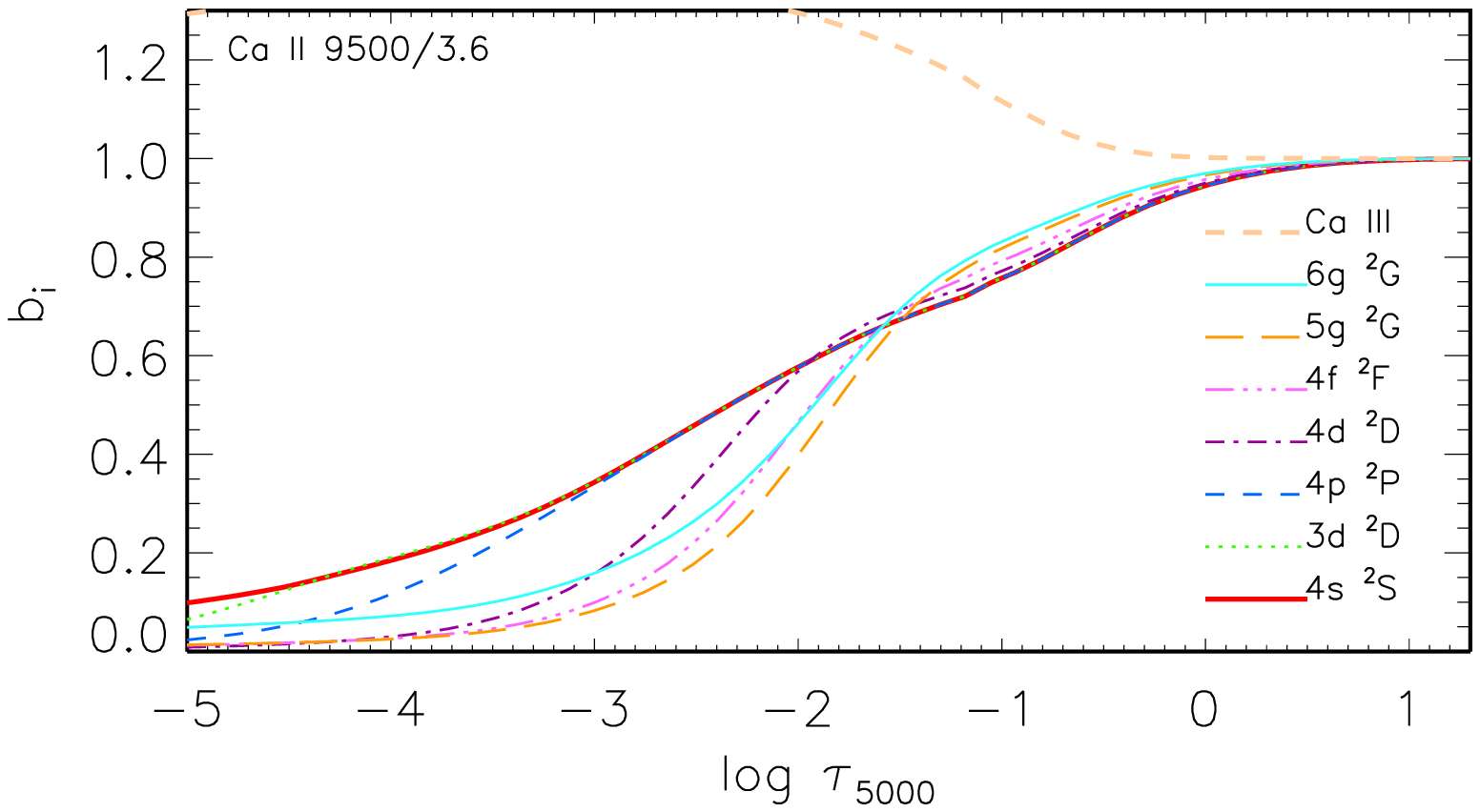}		
	\includegraphics[width=80mm]{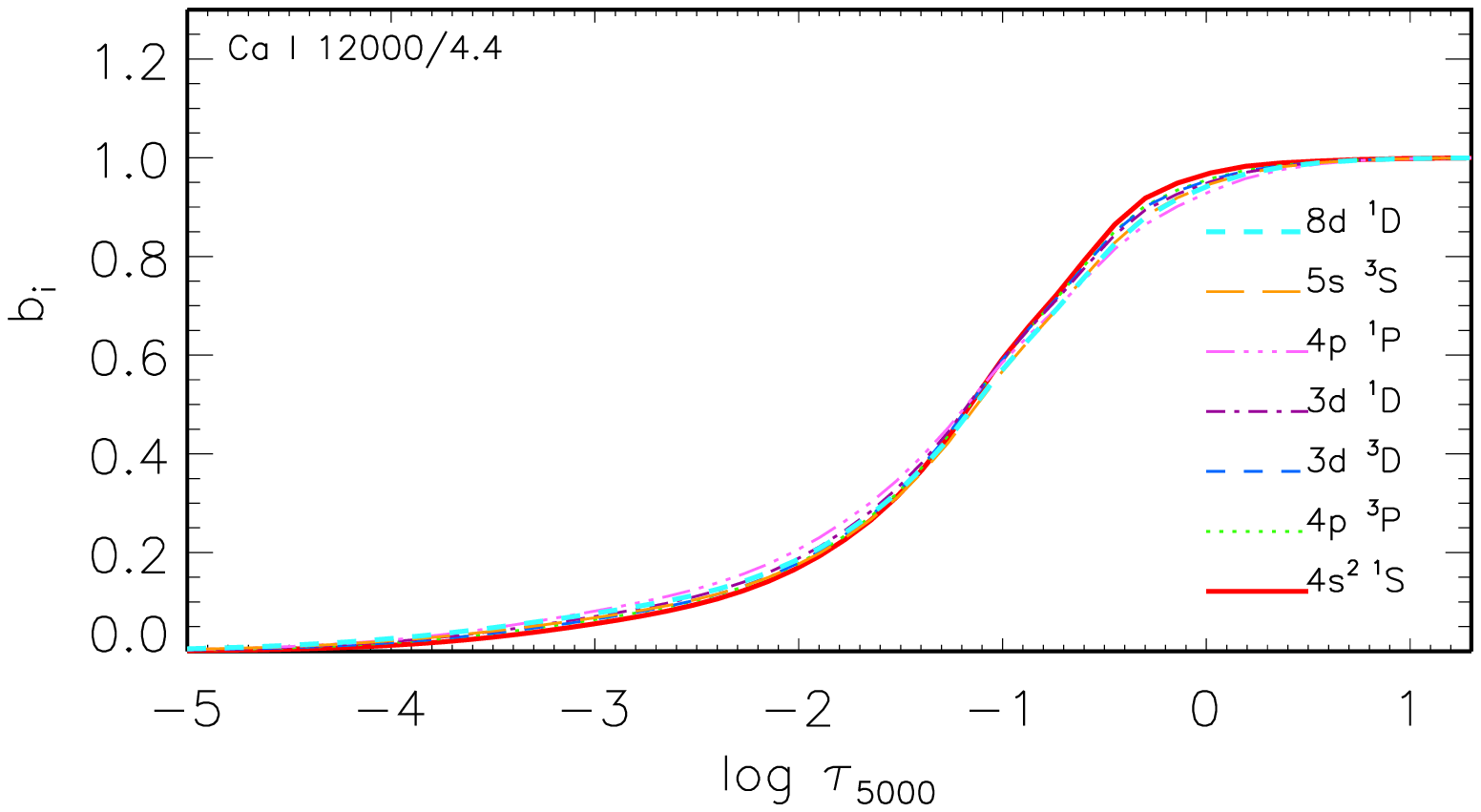}
	\includegraphics[width=80mm]{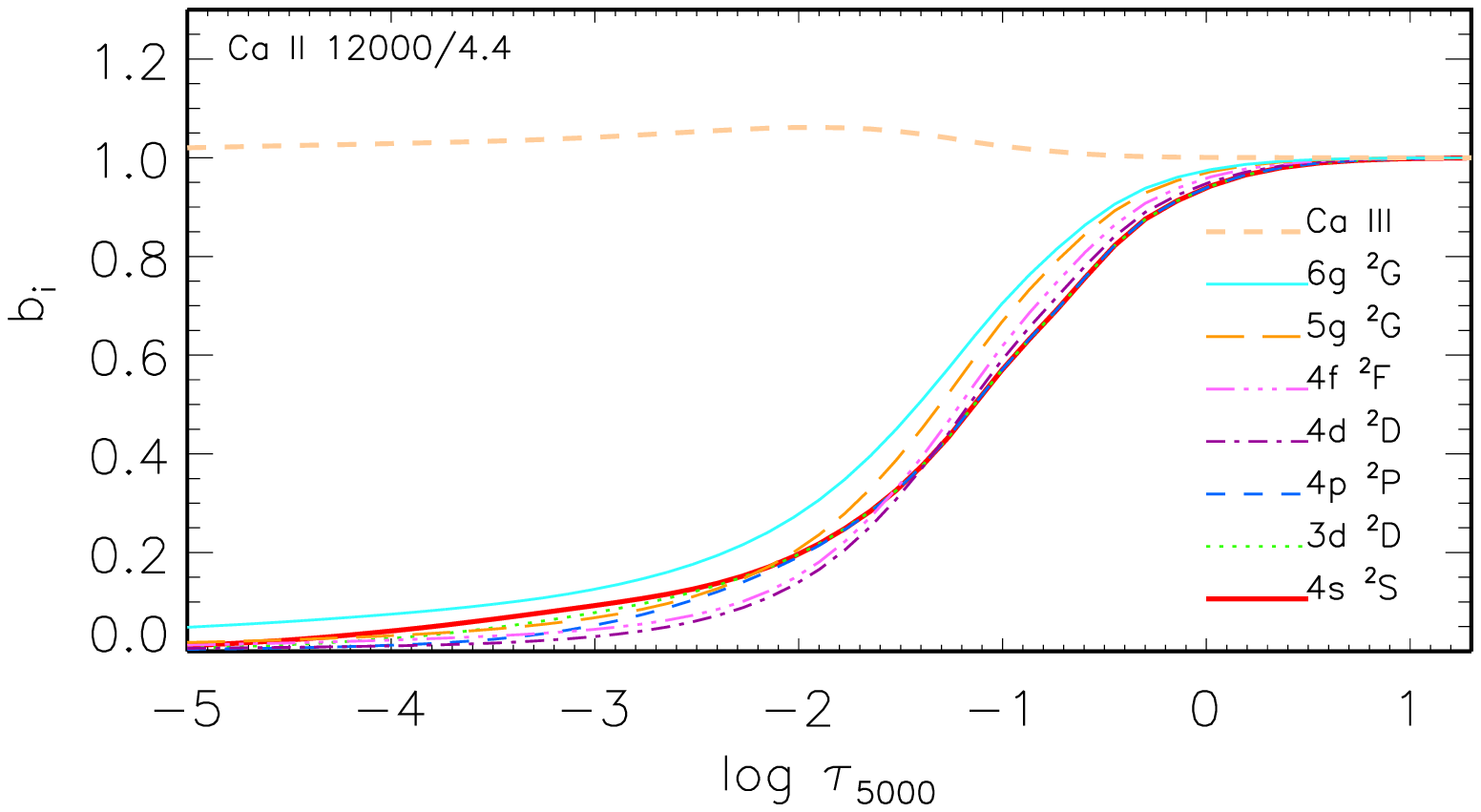}
	\caption{Departure coefficients for the levels of Ca\ione\ (left panels), Ca\ii,   and the ground state of Ca\iii\ (right panels) in different model atmospheres. The selected levels are listed in the order of decreasing \eexc. For each model, the atmospheric parameters \teff / log~g are indicated.  We assume solar chemical composition everywhere.}
	\label{depart} 
\end{figure*}

The deviations from LTE in level populations are characterized by the departure coefficients b$_{i}$~=~n$^{\rm NLTE}_i$/n$^{\rm LTE}_i$, where n$^{\rm NLTE}_i$ and n$^{\rm LTE}_i$ are the statistical equilibrium and thermal (Saha-Boltzmann) number densities, respectively.
The departure coefficients for the selected levels of Ca\ione, Ca\ii, and ground state of Ca\iii\ in different model atmospheres
are presented in Fig.~\ref{depart} as a function of 
log~$\tau_{5000}$.
In each model, all the levels retain their LTE populations in deep atmospheric layers below log~$\tau_{5000}$~=~0, where collisional processes dominate. Deviations from LTE grow toward
higher atmospheric layers. 

The low excitation levels of Ca\ione, with \eexc $\textless$ 4~eV, are underpopulated (b$_i \textless 1$) above log~$\tau_{5000}$~=~0 due to overionisation, caused by superthermal radiation of non-local origin below the ionisation thresholds of these levels. 
In atmospheres with \teff\ ranging from 7000~K up to about 8700~K  (Fig.~\ref{depart}, top left panels), the overionisation of Ca\ione\ decreases toward higher temperature due to an increase of electron number density and collisional rates. 
In atmospheres with 8500~K $\le$ \teff\ $\le$ 10000~K  (Fig.~\ref{depart}, left column, 2nd to 4th rows), the Ca\ione\ 4p$^1$P  level  (dash-three dotted line) is overpopulated relative to the ground state (solid line) at log~$\tau_{5000}$ between 0 and $-1.5$ due to radiative pumping of the resonance 4226  \AA\ transition.

In atmospheres with 7000~K $\le$ \teff\ $\le$ 8000~K, Ca\ii\ is a majority species, and the ground state and  4p$^2$P and 3d$^2$D levels  keep the LTE populations throughout all atmospheric depths. 
The levels of Ca\ii\ with \eexc\ of 6-8~eV (see, for example 4d$^2$D in Fig.~\ref{depart}, top right panel)
are closely coupled to the lower excitation levels inwards at approximately log~$\tau_{5000}$ = $-2$  and depopulated relative to their thermodynamic equilibrium (TE) population in the upper atmospheric layers due to spontaneous transitions. The high-excitation (\eexc\ $>$ 8 eV) levels have depleted populations, starting from deep enough atmospheric layers (for example, log~$\tau_{5000}$ $\simeq$ $-0.5$ in the 7000/4.4 model, Fig.~\ref{depart}).

In atmospheres with \teff\ $\ge$ 10000~K, the levels of Ca\ii\ are underpopulated relative to LTE due to overionisation.
The overionisation is smaller for energy levels with higher \eexc, compared to those, with the lower \eexc, due to collisional coupling of the upper levels with the ground state of Ca\iii. 

We calculated the SE of Ca\ione-\ii\ in the grid of LL model structures \citep{llmod} with: 7000~K $\leq$ \teff\ $\leq$ 13000~K, with a step of 100~K for \teff\ $\leq$ 10000~K and 250~K for \teff\ $\ge$ 10000~K; 3.2 $\leq$ log~g $\leq$ 5.0, with a step of 0.2~dex; $-0.5 \leq$ [Fe/H ] $\leq 0.5$, with a step of 0.5~dex; \vt = 2.0 \kms.
 The grid of departure coefficients can be implemented in any code for NLTE synthetic spectra calculations and is available on request. 

\subsection{NLTE effects for lines of Ca\ione\ and Ca\ii\ depending on atmospheric parameters}

\begin{table}
	\caption{The list of Ca\ione\ and Ca\ii\ lines with the adopted atomic data.}
		\label{atomic}
				\setlength{\tabcolsep}{1.3mm}
		
		\begin{tabular}{|l|c|c|c|c|c|}
			\hline
			$\lambda$ (\AA) & \multicolumn{1}{c}{E$_{exc}$ (eV)} & \multicolumn{1}{c}{ log gf}  & \multicolumn{1}{c}{log $\gamma_{rad}$} & \multicolumn{1}{c}{log $\gamma_4$/N$_e$} & \multicolumn{1}{c}{log  $\gamma_6 $/N$_H$} \\
			\hline
			\multicolumn{6}{l}{Ca\ione} \\
4226.728 &    0.000 &    0.244 &    8.360 &   -6.031 &   -7.562 \\
4283.011 &    1.886 &   -0.136 &    8.320 &   -5.840 &   -7.720 \\
4289.367 &    1.879 &   -0.233 &    8.320 &   -5.760 &   -7.720 \\
4298.988 &    1.886 &   -0.359 &    8.320 &   -5.760 &   -7.720 \\
4302.528 &    1.899 &    0.292 &    8.167 &   -5.997 &   -7.804 \\
4307.744 &    1.886 &   -0.234 &    8.320 &   -5.720 &   -7.720 \\
4318.651 &    1.899 &   -0.139 &    8.320 &   -5.760 &   -7.720 \\
4355.079 &    2.709 &   -0.470 &    7.310 &   -3.550 &   -7.130 \\
4425.437 &    1.879 &   -0.358 &    8.025 &   -5.610 &   -7.163 \\
4434.957 &    1.886 &   -0.007 &    8.021 &   -5.602 &   -7.162 \\
					\hline	
             \multicolumn{6}{p{0.5\textwidth}}{This table is available in its entirety  in a machine-readable form in the online journal. A portion is shown here for guidance  regarding its form and content.}\\	         
			\hline
		\end{tabular}
\end{table}

\label{grid}

\begin{figure}
		\includegraphics[width=90mm]{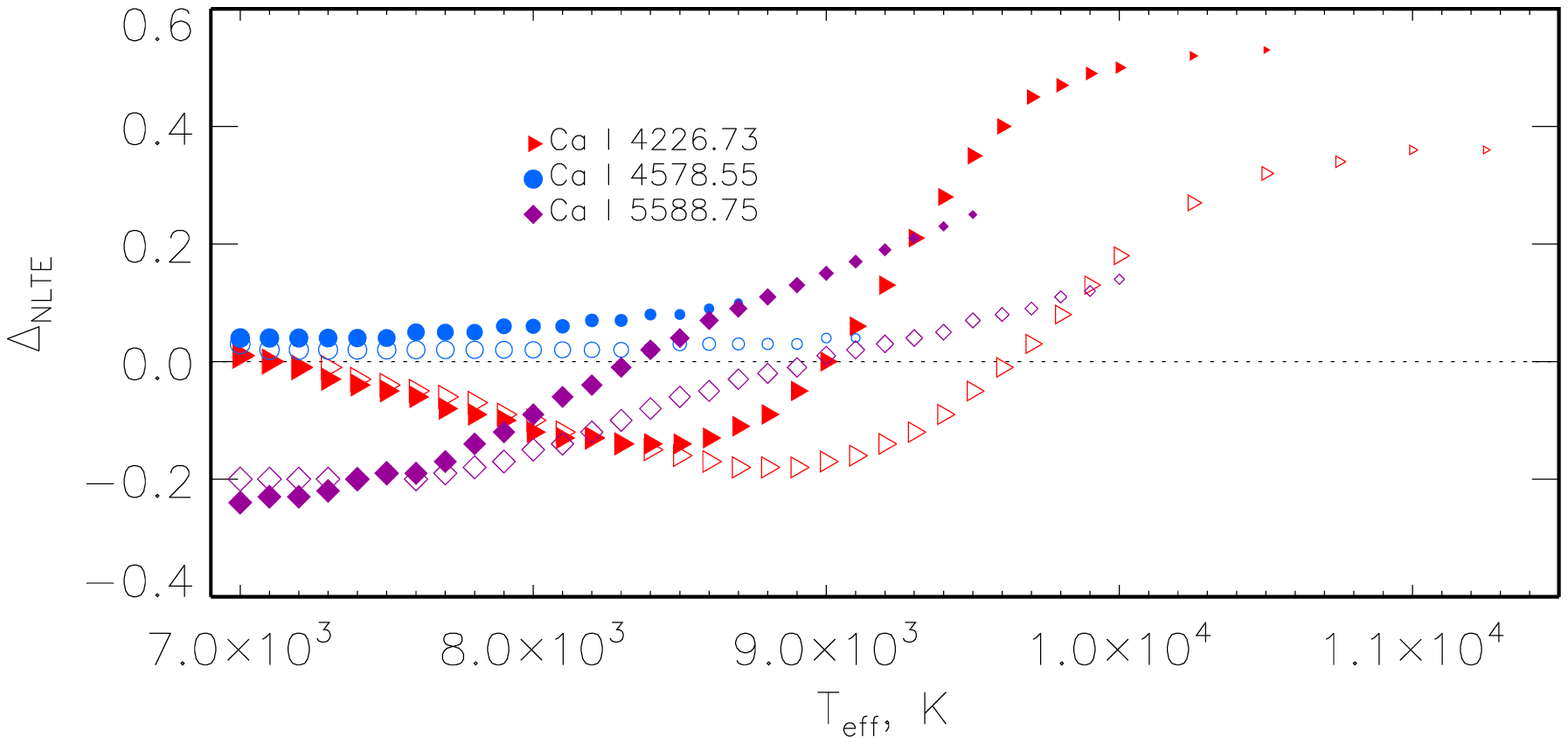}	
		\includegraphics[width=90mm]{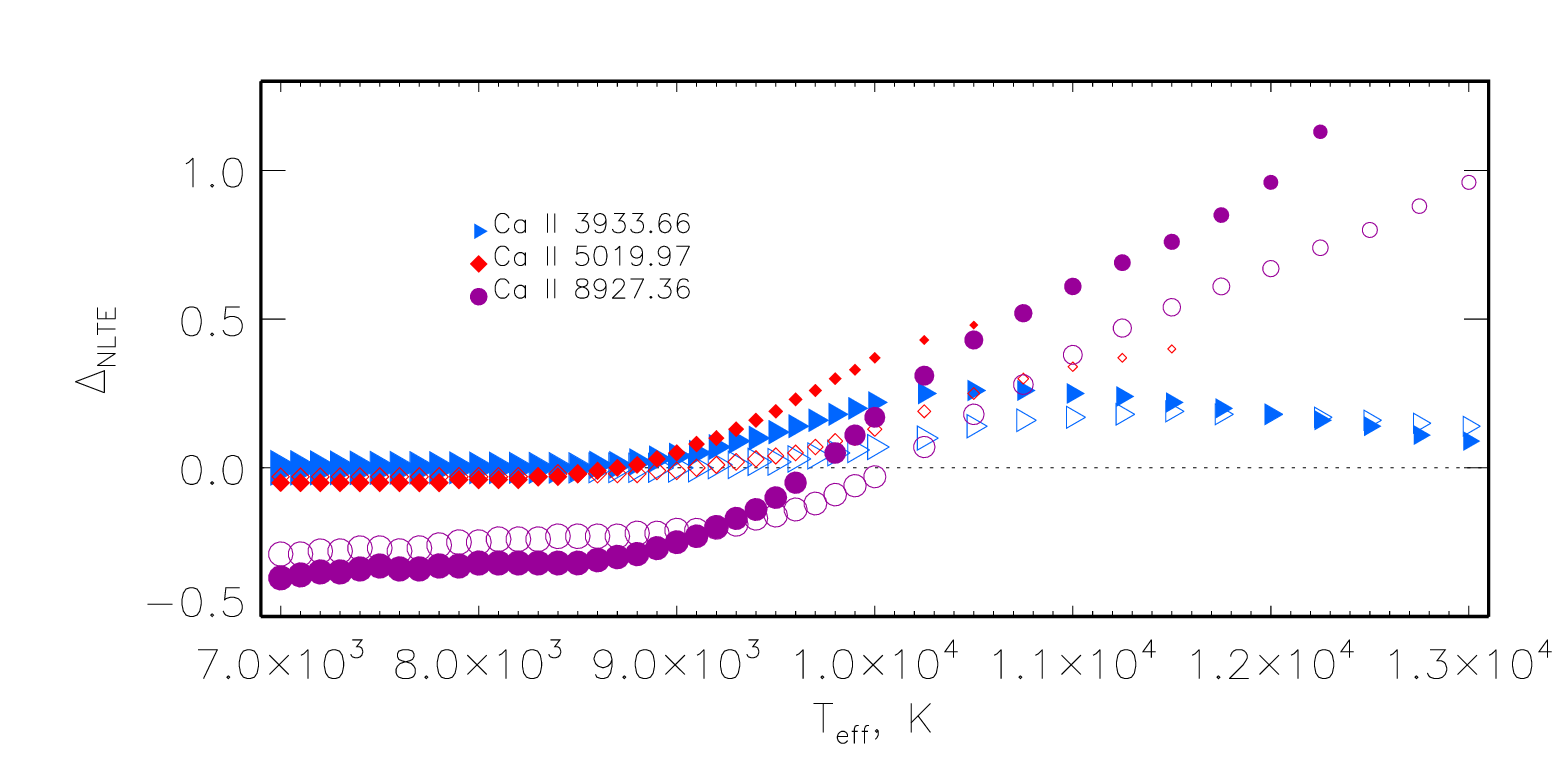}
	\caption{NLTE abundance corrections for the selected lines of Ca\ione\ (top panel) and Ca\ii\ (bottom panel) depending on \teff\ in the models with log~g = 3.6 (filled symbols) and 4.4 (open symbols). 
		 The symbol size is proportional to the line equivalent width.}
	\label{corr}
\end{figure}

The NLTE abundance corrections, $\Delta_{\rm NLTE} = \log A_{\rm NLTE}- \log A_{\rm LTE}$\footnote{Here, log A = log (N$_{elem}$/N$_{tot}$)}, were computed for 55 lines of Ca\ione\ and 13 lines of Ca\ii\ listed in Table~\ref{atomic} along with gf-values, excitation energies, and damping constants (log~$\gamma_{rad}$, log~($\gamma_4$/N$_e$), log~($\gamma_6$/N$_H$). The line list was extracted from the VALD database \citep{2015PhyS...90e4005R}. The NLTE abundance corrections are presented in Table~\ref{corrections} and Fig.~\ref{corr}.

In NLTE, spectral line strength is affected by the deviation of the line source function ($S_\nu$) from the Planck function (B$_{\nu}$) and by the change in the line absorption coefficient ($\chi_\nu$) at line formation depths:

$S_\nu = B_\nu(T) (e^{h\nu_{lu}/kT} - 1)/ (b_l/b_ue^{h\nu_{lu}/kT} - 1)$, 
 
$\chi_\nu \sim b_l(1 - b_u/b_le^{-h\nu_{lu}/kT})$.

\noindent
Here,  b$_l$ and b$_u$ are the departure coefficients of the lower and upper  level of a transition, respectively.
The quantity $b_u/b_le^{-h\nu_{lu}/kT}$ is the correction for stimulated emission that
cancels part of the line absorption.
For the visible and UV lines in the stellar parameter range we are concerned
with, $h\nu_{lu}/kT \gg 1$. Therefore,

$S_\nu \sim B_\nu(T) b_u/b_l$, 

$ \chi_\nu \sim b_l$.

The NLTE abundance corrections for lines of   Ca\ione\ and Ca\ii\  can be either positive or negative, depending on the line and atmospheric parameters.
The weak lines of Ca\ione, such as 4578 \AA, which has an equivalent width of EW $\le$ 50~m\AA\ throughout the investigated range of atmospheric parameters, form in deep atmospheric layers, 
where an overionisation of Ca\ione\ dominates resulting in  positive NLTE abundance corrections.
For  stronger Ca\ione\ lines, 
such as 5588 \AA, 
with EW $\simeq$ 100~m\AA, there are two effects, which influence the NLTE line profile. An overionisation leads to weakened line wings, while photon loss  in high optically thin atmospheric layers strengthens the line core. 
The sign of the NLTE abundance correction depends on which of the two effects prevails at line formation depths.
For 5588 \AA, $\Delta_{\rm NLTE} \le 0$ in atmospheres with \teff~$\le$~8500~K, while  this line is weakened and  $\Delta_{\rm NLTE} \ge 0$ at higher \teff.
Qualitatively similar behaviour is found for  $\Delta_{\rm NLTE}$ for the Ca\ione\ 4226 \AA\ resonance line. However, due to larger equivalent width compared to  5588 \AA, $\Delta_{\rm NLTE (4226)}$ changes the sign at higher effective temperatures, of about 9000~K  when log~g = 4.4.  
 It is worth noting that  \citet{Drake1991} and \citet{mash_ca} found the similar core-wing effect for lines of Ca\ione\  in atmospheres of FG stars.

In atmospheres with 7000~K $\le$ \teff\ $\le$ 8000~K, 
deviations from LTE for Ca\ii\ 3933 \AA\ resonance line and Ca\ii\ IR triplet lines (8498, 8542, 8662 \AA) are small.
NLTE leads to strengthened high-excitation (\eexc $>$ 6~eV)  lines of Ca\ii\ and negative NLTE abundance corrections.
For high-excitation lines of Ca\ii, $\Delta_{\rm NLTE}$ is larger in absolute value for strong  IR lines (8917-27 \AA) compared to weak lines in the visible spectrum range (for example, 5019 \AA, 6456 \AA).

With increasing \teff\ above 9000~K, an overionisation of Ca\ii\ takes place and results in weakened lines of Ca\ii\ and positive $\Delta_{\rm NLTE}$, up to 0.3~dex for Ca\ii\ 3933 \AA\ and 1~dex for 8927 \AA\ (Fig.~\ref{corr}). 

We provide the NLTE abundance corrections for the only	Ca\ii\ 8662~\AA\ line out of the Ca\ii\ IR triplet lines. The Ca\ii\ 8498~\AA\ and 8542~\AA\  lines are  located very close to the Paschen jump, where the hydrogen line profiles calculations for AB-stars can be uncertain  (see Sect.~\ref{ca12_stars}).

The computed NLTE abundance corrections correspond to \vt = 2 \kms.
We checked an effect of change in \vt\ by 1~\kms\ on the results.
The test calculations were performed with the model 7300/4.0/0.0 and \vt~=~1~\kms.
For lines of Ca\ione\ in the 3 m\AA\ $\leq$ EW $\leq$ 227 m\AA\ equivalent width range,  the changes in  $\Delta_{\rm NLTE}$ do not exceed 0.02~dex. For Ca\ii\ lines in the visible spectrum range and of 10 m\AA\ $\leq$ EW $\leq$ 55 m\AA, the changes in  $\Delta_{\rm NLTE}$ are minor and do not exceed  0.01~dex. We found the largest change for Ca\ii\ 8927 \AA\ with EW = 190~m\AA, which amounts to 0.04~dex.

\begin{table}
	\caption{NLTE abundance corrections and equivalent widths for Ca\ione\ and Ca\ii\ lines as a function of  \teff\ and log~g in models with [Fe/H]~=~0 and \vt~=~2~\kms.}
	\begin{large}
		\centering
		\label{corrections}
		\begin{tabular}{|r|r|r|r|r|r|}
			\hline
			log g & \teff$_1$, K & \teff$_2$ & ... &  \teff$_{42}$ & \teff$_{43}$ \\
			\hline	
			& EW$_1$, m\AA & EW$_2$ &  ... &  EW$_{42}$ & EW$_{43}$ \\
			& \dnlte$_1$ & \dnlte$_2$ &  ... &  \dnlte$_{42}$ & \dnlte$_{43}$ \\				
			\hline	
			& 7000 & 7100 &  ... &  12750 & 13000 \\
			\multicolumn{6}{l}{Ca~I    4226.728 \AA,       \eexc =  0.000 eV,    log gf =     0.244 } \\
			3.2 &	258 &    244 &   ... &      --1  &  --1 \\
			&	0.00 & 0.00 &  ... &   --1.00 & --1.00 \\
			\multicolumn{6}{l}{ ... } \\
			5.0 &	109 &    103 &   ... &      --1  &  --1 \\
			&	0.01 & 0.01 &  ... &   --1.00 & --1.00 \\
			\multicolumn{6}{l}{ ... } \\
			\multicolumn{6}{l}{Ca~II    9890.705 \AA,       \eexc =  8.400 eV,    log gf =    -1.200} \\
			3.2 &   201 &    206 &   ... &   --1 &     --1 \\
			& --0.30 &  --0.30 &  ... &   --1.00 & --1.00	\\
			\multicolumn{6}{l}{ ... } \\
			5.0 &  172 &    164 &   ... &  13 &     10 \\
			& --0.08 &  --0.08 &  ... &   1.91 & 2.33 	\\	
			\hline	
			\multicolumn{6}{p{.47\textwidth}}{This table is available in its entirety in a machine-readable form in the online journal. A portion is shown here for guidance regarding its form and content. If EW~=~$-1$ and \dnlte~=~$-1$, this means that EW~$<$~5~m\AA\ in a given model atmosphere. 
} \\
			\hline
		\end{tabular}
	\end{large}
\end{table}

\section{Calcium NLTE abundances of the reference stars}
\label{hots}

As a test of the NLTE method for Ca\ione-\ii\ in atmospheres with \teff $>$ 7000~K, we determine calcium abundances  in the atmospheres  of nine well-studied sharp-lined stars of B-F spectral types, which do not reveal the presence of pulsation, chemical stratification, and magnetic field. These stars were used  in our previous studies for testing the NLTE methods for  C\ione-\ii\ \citep{Alexeeva_c_hot}, O\ione\ \citep{sitnova_o}, and Ti\ione-\ii\  \citep{Sitnova_ti}.
 The list of program stars  together with their atmospheric parameters is given in Table~\ref{obs}.
 We refer to \citet{Alexeeva_c_hot} and \citet{Sitnova_ti} for a description of the methods of atmospheric parameter determinations. 

\subsection{Observations }

We use high-resolution, $\lambda/\Delta\lambda \ge$~40000, and high signal to noise ratio, S/N $\ge$~100,  spectra with wide wavelength coverage, allowing us to use Ca\ii\ 3933 \AA\ and 9890 \AA\ lines. The exception is HD~145788, where only the 3780-6910 \AA\ wavelength region is available.
 Details of the observations and of their reduction are given in the references (last column of Table~\ref{obs}).

Inspecting the observed spectrum of 21~Peg, we found additional absorption in the red wings of Ca\ii\ 3933~\AA\ and 3968~\AA\ lines and the Na\ione\ 5889~\AA\ and 5895~\AA\ lines.
These features have a similar radial velocity shift of 12 \kms, with respect to the stellar spectrum, suggesting an interstellar origin. 

 It is not surprising that a nearby star 21~Peg (at a distance of 190 pc from the Sun and E(B-V) = 0.04) presents interstellar medium absorption lines.
\citet{Fossati2017} concluded that the core of the Ca\ii\ H\&K lines in spectra of stars lying on average beyond 100 pc may be significantly affected by interstellar medium absorption.
To derive calcium abundance from Ca\ii\ 3933~\AA, we use spectral fitting of the blue wing of the line.

\begin{table*}
	\caption{Stellar atmospheric parameters and characteristics of the observed spectra.}
	\begin{large}
		\centering
		\label{obs}

		\begin{tabular}{lcccccccc}
			\hline
			Star & \teff ,  K & $\log g$ & [Fe/H] & \vt, \kms  &   Ref. &$\lambda/\Delta\lambda$, 10$^3$  & $S/N >$ & source  \\
			\hline
			HD~32115 &  7250 & 4.20 &   0.0 & 2.3 & F11  & 60 & 490 &  CAO  \\
HD~73666 & 9380 & 3.78 & 0.10 & 1.8 & F07  & 65 & 660 &  F07 \\
HD~172167  & 9550 & 3.95 & -0.50 & 1.8 & CK93  & 40 & 750 & K99  \\
(Vega) &     &    &  &&&&& \\
			HD~72660  & 9700 & 4.10 & 0.40 & 1.8 & S16  & 65 & 100 & K18  \\ 
			HD~145788 & 9750 & 3.70 & 0.46 & 1.3 & F09 & 115 & 200 &  F09  \\
			HD~48915 & 9850 & 4.30 &   0.40 & 1.8$^1$ & H93  & 70 & 500 & F95  \\
			(Sirius) &     &    &   &&&&&                  \\
			HD~209459 & 10400 & 3.55 & 0.0 & 0.5 &  F09 & 120 & 700 &  CAO  \\
			(21 Peg) &     &    &  &&&&&                  \\
			HD~17081 & 12800 & 3.75 & 0.0 & 1.0 &   F09 & 65 & 200 &  F09  \\
			($\pi$ Cet) &     &    &         &&&&&            \\
			HD~160762 & 17500 & 3.8 & 0.02 & 1.0 &   NP12 & 65  & 600  & CAO  \\
			($\iota$ Her) &     &    &         &&&&&            \\
			\hline
             \multicolumn{9}{p{1.0\textwidth}}{
            $^1$  \cite{Sitnova2013}; 
F11 = \cite{fossati11}, S16 = \cite{Sitnova_ti}, CK93 = \cite{1993ASPC...44..496C}, 
F07 = \cite{fossati07}, F09 = \cite{fossati09}, H93 = \cite{hill1993}, NP12 = \cite{NP12}, K18 = \cite{khalack2018},  ESPaDOnS, Propisal ID = 1896062;   K99 = A.~Korn, private communication, FOCES; 
F95 = \cite{Furenlid1995}, L98 = \cite{Landstreet1998}, CAO = taken from the Common Archive Observation database http://www.cfht.hawaii.edu/Instruments/Spectroscopy/Espadons/.}\\	
			\hline			
		\end{tabular}
	\end{large}
\end{table*}

\subsection{Analysis of Ca\ione\ and Ca\ii\ lines in the reference stars}
\label{ca12_stars}

Here we determine NLTE and LTE abundances from different lines of Ca\ione\ and Ca\ii\ and check the Ca\ione--\ii\ ionisation equilibrium in the investigated stars.  
 We emphasize that  the  abundances  are  derived  from the spectrum synthesis and not from equivalent widths.
For each star, at least, six calcium lines were detected.
We present the derived NLTE and LTE abundances  obtained from individual lines of Ca\ione\ and Ca\ii\ in Table~\ref{linebyline}.
The lines of two ionisation stages are observed in  seven stars out of nine.
We separated the lines into four groups: Ca\ione\ lines, Ca\ii\ lines in the visible spectrum range (3933, 4220, 5001, 5019, 5021, 5285, 5307, 5339, 6456 \AA), Ca\ii\ high excitation  IR lines (4d~$^2$D--4f~$^2$F, 8912 \AA\ and 8927 \AA\ and 4f~$^2$F--5g~$^2$G,  9890~\AA), and Ca\ii\ IR triplet lines (3d~$^2$D--4p~$^2$P, 8498, 8562, 8662 \AA). 
Further, we discuss separately the abundances, derived from each group of lines in the investigated stars.

\begin{figure}
	\includegraphics[width=80mm]{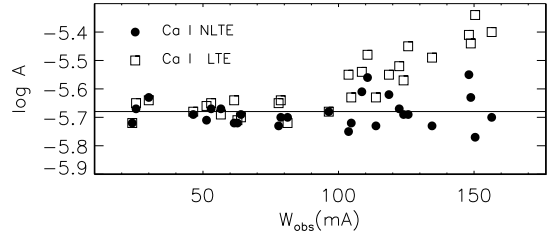}
	\caption{NLTE (filled circles) and LTE (open squares) abundances from Ca\ione\ lines in HD~32115 depending on the line equivalent width. The average NLTE abundance from Ca\ione\ is shown with solid line. }
	\label{ca1_hd32115}
\end{figure}

\underline{Ca\ione\ lines.} In the five A-type stars, namely HD~73666, Vega, HD~72660, HD~145788, and Siruis, the  Ca\ione\ 4226 \AA\ resonance line and the subordinate lines 
were  used for abundance determination. In the F-type star, HD~32115, Ca\ione\ 4226 \AA\ was excluded from the abundance analysis due to its large strength (EW~$\ge 330$ m\AA) and blending with Fe\ione\ 4226.424~\AA\ line. 
For A-type stars, NLTE leads to higher calcium abundance compared to LTE, by up to 0.33~dex in HD~145788.
For HD~32115, the NLTE abundance corrections  for different lines of Ca\ione\ have different sign (Fig.~\ref{ca1_hd32115}), resulting in 0.11~dex lower average abundance compared to the LTE case. 
In A-type stars, the NLTE abundance correction for Ca\ione\ 4226~\AA\ is larger compared to that for the  Ca\ione\ subordinate lines.
For example, in HD~145788 $\Delta_{\rm NLTE, 4226} = 0.44$~dex, while for the subordinate lines, $\Delta_{\rm NLTE}$ varies from 0.27~dex to 0.29~dex. 
For Ca\ione\ in each star, NLTE leads to smaller line-to-line abundance scatter compared to LTE (see Table~\ref{abund}).

\underline{Ca\ii\ lines in the visible spectrum range.} NLTE leads to higher calcium abundance compared to LTE, by up to 0.35~dex in 21~Peg. 
The exception is the coolest star of our sample, HD~32115, where the NLTE abundance corrections are either negative or minor for lines of Ca\ii. 
NLTE leads to smaller line-to-line abundance scatter compared to LTE in each star.

\underline{Ca\ii\ high excitation  IR lines.} These lines are the most sensitive to deviations from LTE.
The NLTE abundance corrections are negative for the stars with the strongest lines, HD~32115, HD~73666, and HD~72660, and positive for the other stars of the sample. 
With increasing \teff, the Ca\ii\  IR lines weaken and form in atmospheric layers, where overionisation dominates, which leads to positive NLTE abundance corrections. 

The Ca\ii\ 8912-27 and 9890 \AA\ lines were not detected in the observed spectrum of HD~17081. 
Our NLTE calculations with the NLTE abundance derived from Ca\ii\ 3933 \AA\ predict a tiny absorption in the  IR lines, which is consistent with the observations, while LTE leads to strong absorption.  The difference between LTE and NLTE abundances amounts to 0.8 dex.

\begin{figure}
	\includegraphics[width=80mm]{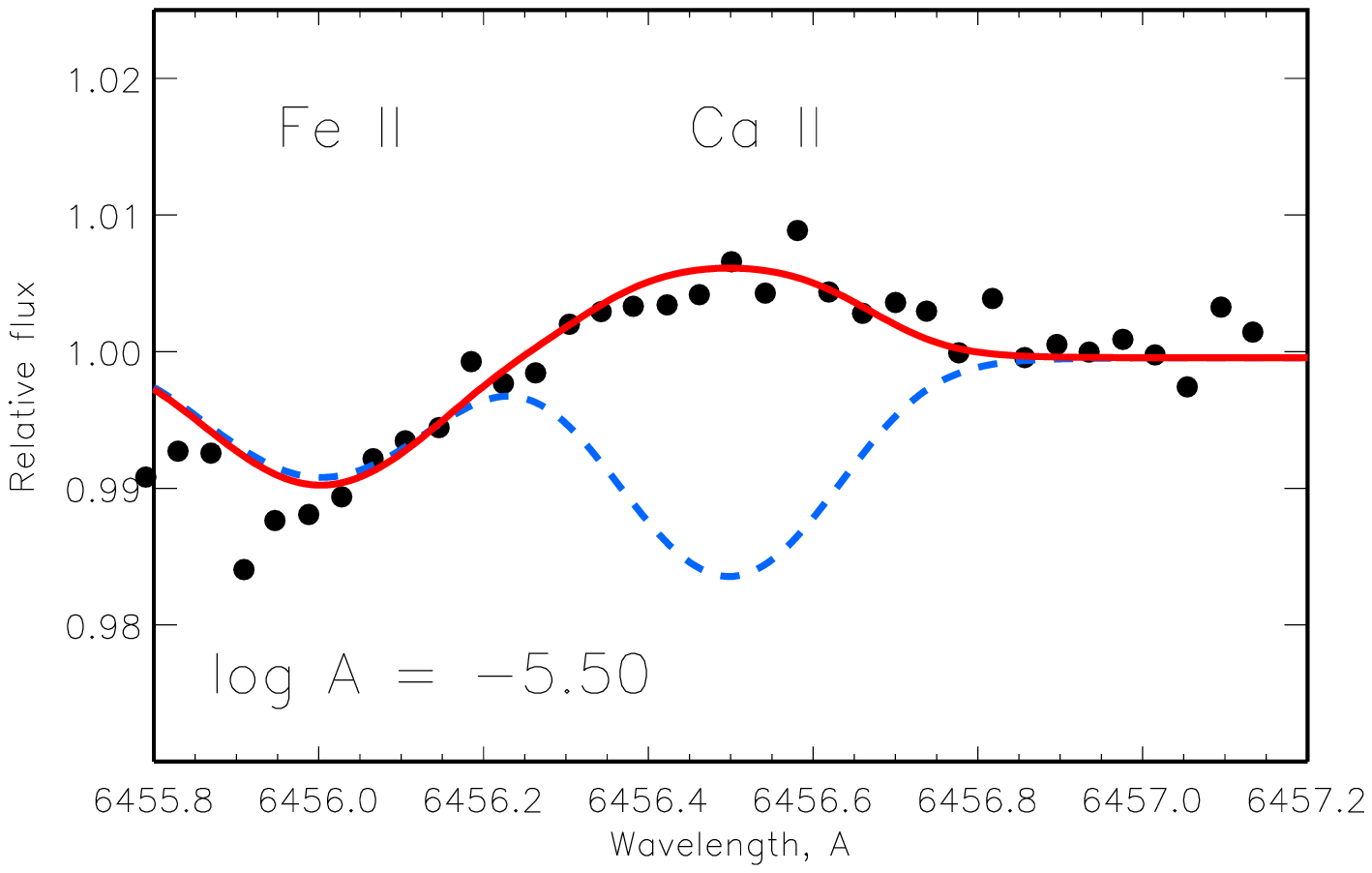}
	\includegraphics[width=80mm]{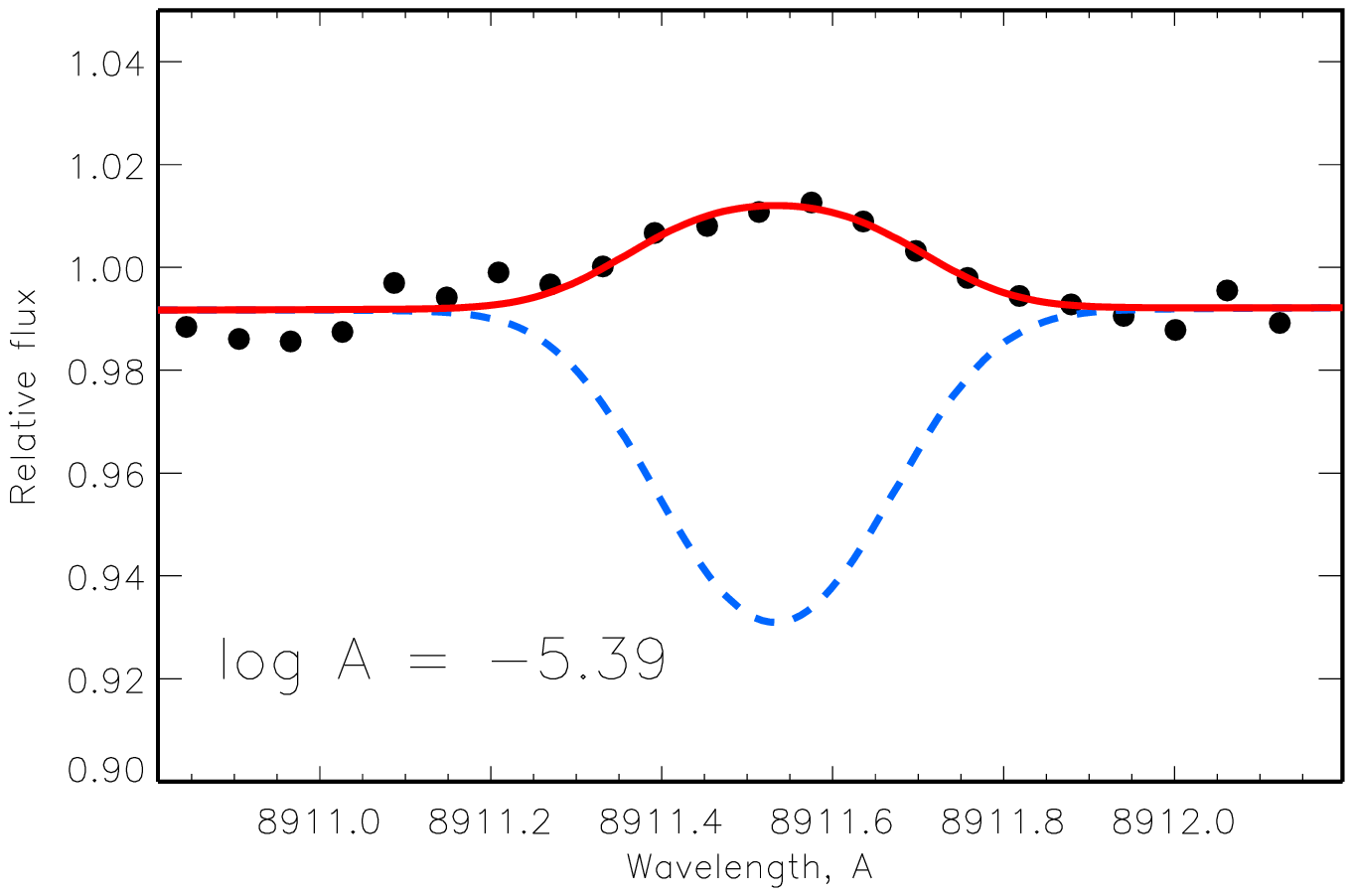}
	\includegraphics[width=80mm]{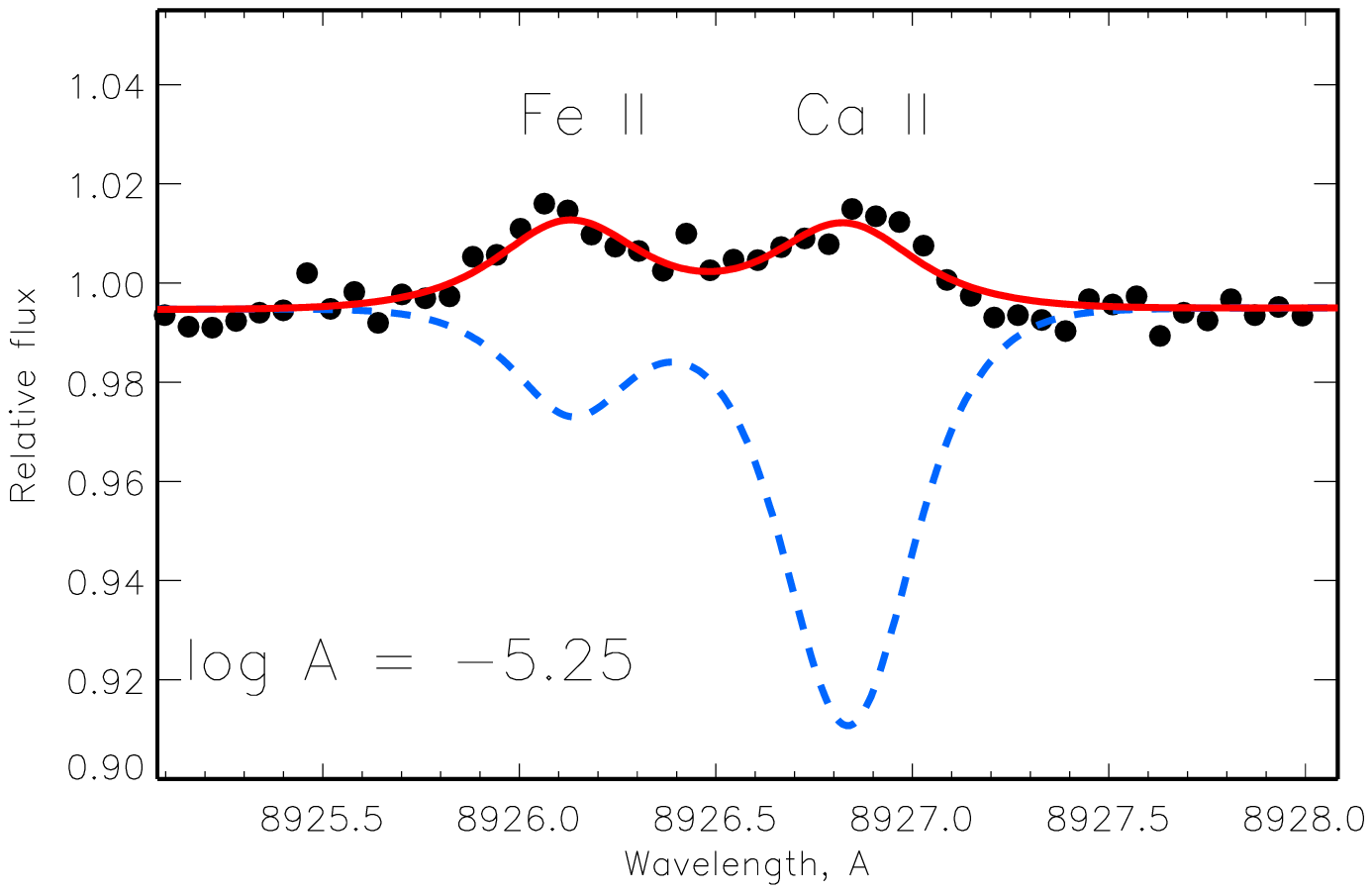}
	\includegraphics[width=80mm]{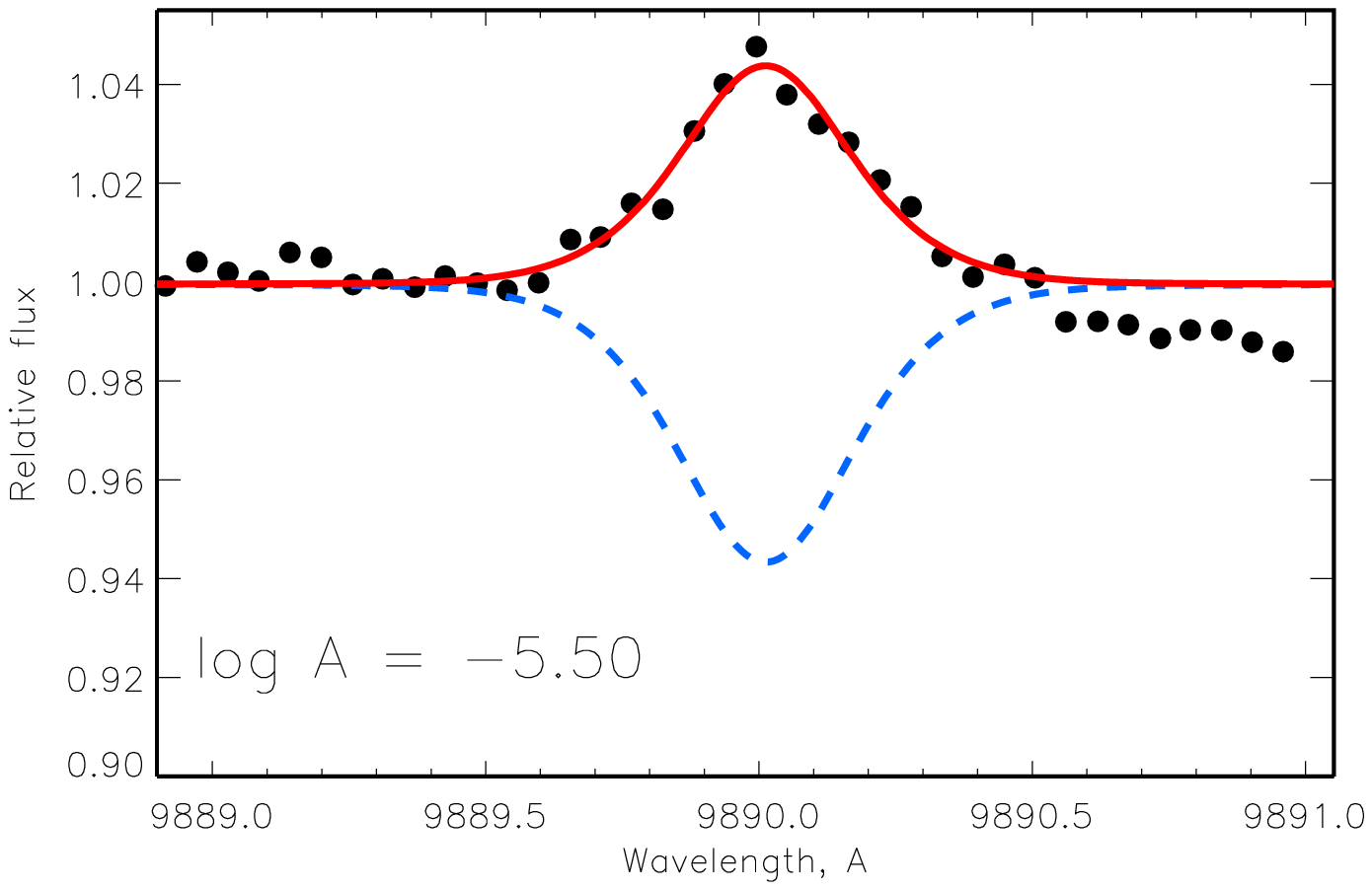}
	\caption{NLTE (solid curves) and LTE (dashed) line profiles of Ca\ii\ 6456, 8912, 8927, and 9890 \AA\ in HD~160762. The best fit calcium NLTE abundance is indicated for each line. 
In the first and the third panels, the NLTE profiles are shown  also considering the Fe\ii\ 6456 \AA\ and 8926 \AA\ lines blending the Ca lines.
		The NLTE calculations for Fe\ione-\ii\ will be presented in an upcoming paper. }
	\label{emission}
\end{figure}

\begin{figure}
		\includegraphics[width=80mm]{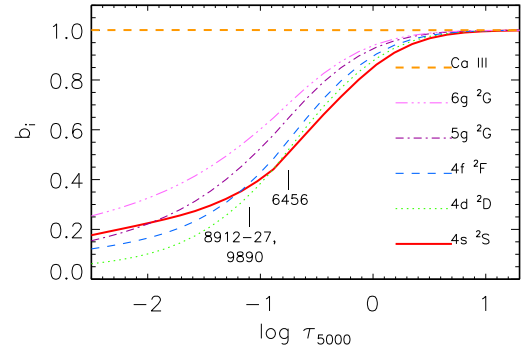}
	\caption{Departure coefficients for the selected levels of Ca\ii\   and the ground state of Ca\iii\  in the model atmosphere of HD~160762, with \teff\ = 17500~K and log~g = 3.8. The Ca\ii\ 6456, 8912-27, and 9890 \AA\ lines form  in the 4f~$^2$F--6g~$^2$G, 4d~$^2$D--4f~$^2$F and 4f~$^2$F--5g~$^2$G transitions, respectively. Tick marks indicate the locations of line core formation depths for the quoted Ca\ii\ lines.
}
	\label{b_emission}
\end{figure}

\begin{figure}
	\includegraphics[width=80mm]{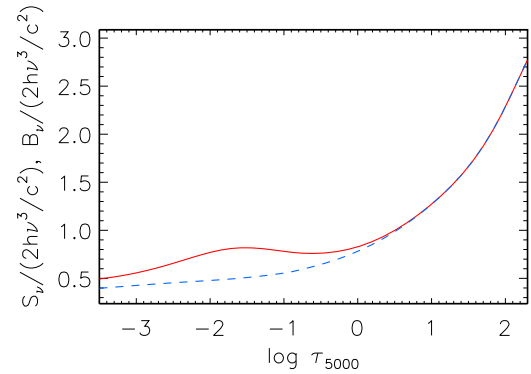}
	\caption{The Planck function (dashed curve) and the line source function (solid curve) in dimensionless units for the Ca\ii\ 9890~\AA\ line. The line core forms at log~$\tau_{5000}$ = $-1.1$. }
	\label{sn_bn}
\end{figure}

\underline{The emission lines of Ca\ii\ in HD~160762.}
In the hottest star of our sample, we found out that the Ca\ii\ 6456, 8912, 8927, and 9890 \AA\ lines are observed in emission.
There is no way to reproduce emission lines using classical hydrostatic model atmosphere and  LTE level populations.
Our NLTE calculations with the hydrostatic LTE model atmosphere and without any tuning, reproduce well the observed Ca\ii\ 6456, 8912, 8927, and 9890 \AA\ line profiles, as displayed in Fig.~\ref{emission}.
To understand the emission mechanism,
we show in Fig.~\ref{b_emission} the departure coefficients of some important Ca\ii\ levels. All of them are underpopulated compared to their LTE populations, with higher
\eexc\ levels closer to LTE, which is typical for the levels of a
minority species affected by overionisation and photon loss
in high-excitation lines. As a result, for each of the Ca\ii\ 6456, 8912-27 and 9890 \AA\ lines, their line source function, S$_\nu$, exceeds the Planck function, B$_\nu$(T$_e$), at the line formation depths. 
It is worth noting that the strength of the emission lines increases with wavelength. For example, the equivalent width of the Ca\ii\ 6456~\AA\ line is EW$_{6456}$ = $-2$~m\AA, while EW$_{9890}$ = $-18$~m\AA\ for the Ca\ii\ 9890~\AA\ line.
In the IR region, deviations from LTE are larger
compared to the visible spectral range due to the increasing
importance of the stimulated emission, which also increases
with temperature (see the relation in Sect.~\ref{se}).
Figure~\ref{sn_bn} displays S$_\nu$ for Ca\ii\ 9890~\AA\ and the corresponding B$_\nu$(T$_e$) in the atmosphere of HD~160762. The core of the Ca\ii\ 9890~\AA\ line forms around log~$\tau_{5000}$ = $-1.1$, where S$_\nu$ exceeds B$_\nu$(T$_e$) by a factor of 1.6.
Note that the population divergence does not reach the regime of ''laser action'' (negative line extinction and negative line source function),  which would require b$_u$/b$_l$ $>$ 2.7 and 4.7 for the 9890~\AA\ and 6456~\AA\ lines, respectively, when T = 14360~K and log~$\tau_{5000}$ = $-0.9$. The  computed divergence remains below this value, and b$_u$/b$_l$ $<$ 2 for any emission line.

\underline{Ca\ii\ 8248-54 \AA\ and the Ca\ii\ IR triplet lines}  were analyzed in the eight stars of the sample, where the near IR observed spectra are available. 
We found consistent within 0.14~dex NLTE abundances from Ca\ii\ lines in the visible spectrum range, the IR triplet, and 8248-54 \AA\ lines in HD~32115, HD~73666, and Sirius. For HD~72660 and HD~17081, Ca\ii\ 8248-54 \AA\ lines are too weak to be detected, while NLTE abundances from the IR triplet and the other Ca\ii\ lines are consistent within 0.09~dex. 
For Vega, 21~Peg, and HD~160762, different groups of Ca\ii\ lines give different abundances either in NLTE or LTE, by up to 0.2~dex in absolute value. 
For example,  the  Ca\ii\ IR triplet lines give higher abundances compared to Ca\ii\ lines in the visible spectrum range in Vega and 21~Peg, while lower in HD~160762. Abundances from Ca\ii\ 8248-54 \AA\ are higher, lower, and consistent with those derived from Ca\ii\ visible lines in HD~160762, 21~Peg, and Vega, respectively.
The uncertainty in abundance from Ca\ii\ 8248-54 \AA\ and the Ca\ii\ IR triplet lines can be caused by the following reasons.

\begin{figure}
	\includegraphics[width=80mm]{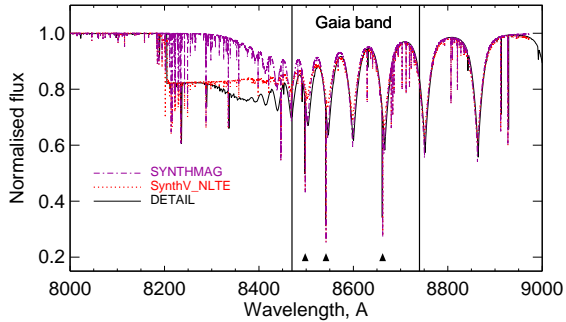}
	\caption{
		Synthetic spectra in the \teff\ = 10400 K and log g = 3.55 model atmosphere calculated with different treatments of the hydrogen-line opacity, using the codes DETAIL (solid curve), SynthV\_NLTE (dotted curve), and SYNTHMAG (dash-dotted curve). The position of the Ca\ii\ IR triplet lines is marked with triangles. The vertical lines indicate the wavelength range covered by the Gaia spectra.
	}
	\label{paschen}
\end{figure}

The Ca\ii\ 8248-54 \AA\ lines lie near the Paschen jump and the Ca\ii\ triplet lines lie near the cores of the hydrogen Paschen lines. 
This is not a trivial problem to normalise an observed spectrum in the wavelength region from 8200~\AA\ to 8800~\AA.
To fit the theoretical  spectrum to the observed one, we have to additionally correct a local continuum for  individual IR lines.
Moreover, in this wavelength region, there are uncertainties in the calculations of the theoretical spectrum for BA-type stars.
Figure~\ref{paschen} shows the synthetic spectra 
calculated with different treatments of the hydrogen-line opacity in the model atmosphere with \teff\ /log~g = 10400/3.55 representing 21~Peg.
The calculations were performed  with the  codes  DETAIL \citep{detail} and  SynthV\_NLTE \citep{Tsymbal1996,ryab2015}. 
Additionally, we calculated the synthetic spectrum with the SYNTHMAG code \citep{Kochukhov2007}, 
which uses  HBOP procedure, developed by  \citet{2015ascl.soft07008B}, for treatment of
the overlapping hydrogen lines
and continuous opacity with
the occupation probability theory  \citep{Hummer1988,Dappen1987,Hubeny1994,Nayfonov1999}.
The use of the occupation probability formalism significantly  reduces Paschen line absorption compared to those from classical treatment of the hydrogen line opacity. 
Therefore, we do not  provide abundance measurements from these lines in the investigated stars and do not recommend to use Ca\ii\  8248-54 \AA\ and the Ca\ii\ IR triplet lines for abundance determinations for BA type stars.

\begin{figure}
	\includegraphics[width=80mm]{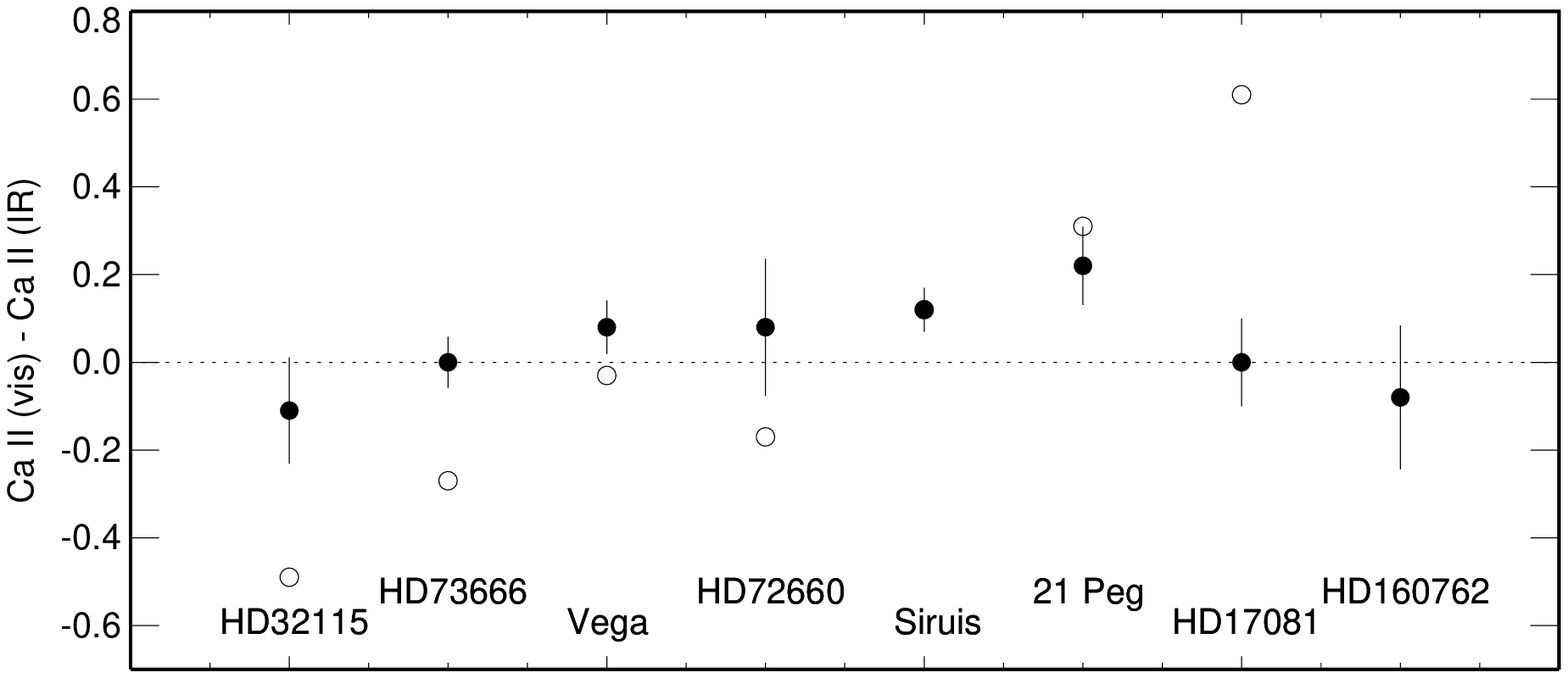}	
	\includegraphics[width=80mm]{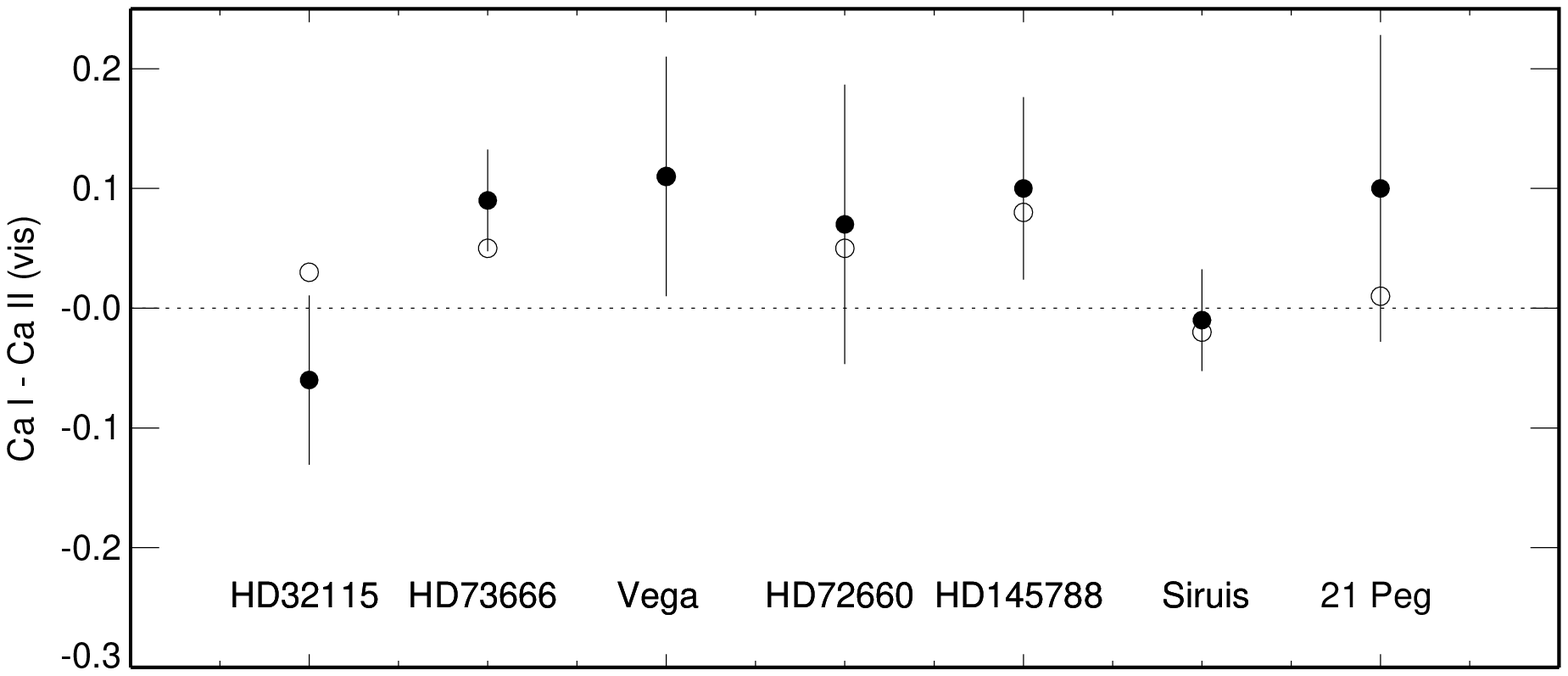}
	\caption{
NLTE (filled circles) and LTE (open circles) abundance differences between Ca\ii\  lines in the visible and  IR spectral range (top panel) and between Ca\ione\ lines and Ca\ii\ lines in the visible spectral range (bottom panel) in the investigated stars.
		}
	\label{abund_stars} 
\end{figure}

 Table~\ref{abund} summarises the results of calcium abundance determination from different groups of  reliable lines, i. e.  lines of Ca\ione,  lines of Ca\ii\ in the visible spectrum range, and  lines of Ca\ii\ in the  IR spectrum range 8912-27, 9890 \AA. We also show in Fig.~\ref{abund_stars} the NLTE and LTE abundance differences between these groups of lines.
For each star, NLTE leads to average abundances from lines of Ca\ii\ in the visible and the  IR spectrum range consistent within 0.1~dex, while the abundance discrepancy between these groups of lines varies from 0.6~dex down to  $-0.5$~dex in LTE (Fig.~\ref{abund_stars}, top panel). 
The exception is 21~Peg, where NLTE leads to better agreement between these two groups of lines compared to LTE, however, a  discrepancy of 0.2~dex remains.
For HD~160762, the LTE abundance difference is not shown in Fig.~\ref{abund_stars}, because Ca\ii\ 8912-27, 9890 \AA\ lines are observed in emission and can not be fitted in LTE. 

Fig.~\ref{abund_stars}, bottom panel, shows the NLTE and LTE abundance differences between lines of Ca\ione\ and Ca\ii\ in the visible spectrum range. 
For the majority of the investigated stars, an abundance  difference $\Delta_{\rm CaI-CaII}$ = log~A(Ca\ione)--log~A(Ca\ii, vis) does not exceed 0.1~dex in absolute value, independent of  NLTE or LTE. 
However, for A-type stars, the absolute abundances from Ca\ione\ and Ca\ii\ lines are higher in NLTE compared with LTE.

\begin{table*}
	\caption{
		NLTE and LTE abundances from individual  Ca\ione\ and Ca\ii\ lines in the reference stars.
	 Here, log~A(X) = log($N_X/N_{\rm tot}$).}
	\renewcommand\arraystretch{1.1}
	\centering
	\label{linebyline}
	
	\begin{tabular}{|l|r|r|r|r|r|r|r|r|r|r|r|r|r|r|}
		\hline
		&  &  & \multicolumn{2}{c}{HD 32115}	 & \multicolumn{2}{c}{HD~73666 }	&\multicolumn{2}{c}{Vega}	& \multicolumn{2}{c}{HD 72660}	 & \multicolumn{2}{c}{HD 145788} & ... \\
		&  &  & \multicolumn{2}{c}{log~A}	 & \multicolumn{2}{c}{log~A }	&\multicolumn{2}{c}{log~A}	& \multicolumn{2}{c}{log~A}	 & \multicolumn{2}{c}{log~A} &  \\
		$\lambda$, \AA & \eexc, eV & log gf & NLTE & LTE	 & NLTE & LTE	& NLTE & LTE	& NLTE & LTE	 & NLTE  & LTE &  \\
		\hline
		4226.728 &    0.000 &  0.244 & -99.9  & -99.9  &   -5.50 &  -5.58 &   -6.03 &  -6.39 &   -5.43 &  -5.47 &   -5.26 &  -5.70 & ... \\
		4302.528 &    1.899 &  0.292 & -5.63 & -5.44  &   -5.53 &  -5.66 & -99.9   & -99.9  &   -5.47 &  -5.56 & -99.9   & -99.9 & ...\\
		4425.437 &    1.879 & -0.358 &  -5.62 &  -5.55 &   -5.53 &  -5.65 & -99.9   & -99.9  &   -5.44 &  -5.54 & -99.9   & -99.9 & ...\\
		\hline
		\multicolumn{15}{p{.947\textwidth}}{This table is available in its entirety in a machine-readable form in the online journal. A portion is shown here for guidance regarding its form and content. If log~A~=~$-99.9$, this means that a line was not used for abundance determination.} \\
	\end{tabular}
\end{table*}

\begin{table*}
	\caption{
Average NLTE and LTE abundances from  Ca\ione\ lines, Ca\ii\ lines in the visible spectrum region (Ca\ii, vis), and Ca\ii\ lines in the  IR spectrum region (Ca\ii, IR includes 8912-27, 9890  \AA) in the investigated stars.
}
	\renewcommand\arraystretch{1.1}
		\centering
		\label{abund}

		\begin{tabular}{|l|r|l|c|l|r|c|l|c|}
			\hline
Star      & & log A(Ca\ione) &  N$_{\rm Ca I}$ & log A(Ca\ii, vis) & [Ca/H](Ca\ii, vis) &  N$_{\rm Ca II, vis}$   & log A(Ca\ii, IR) & N$_{\rm Ca II, IR}$ \\
			\hline
HD~32115 & NLTE &   -5.68 $\pm$   0.05 &  22 &  -5.62 $\pm$   0.05 &  0.09 &   5 &  -5.51 $\pm$   0.11 &   3 \\                       
&  LTE &   -5.57 $\pm$   0.13 &  22 &  -5.60 $\pm$   0.06 &  0.11 &   5 &  -5.11 $\pm$   0.20 &   3 \\                         
HD~73666 & NLTE &   -5.48 $\pm$   0.03 &  11 &  -5.57 $\pm$   0.03 &  0.14 &   7 &  -5.57 $\pm$   0.05 &   3 \\                       
&  LTE &   -5.59 $\pm$   0.06 &  11 &  -5.64 $\pm$   0.03 &  0.07 &   7 &  -5.37 $\pm$   0.09 &   3 \\                       
Vega     & NLTE &   -5.94 $\pm$   0.08 &   5 &  -6.05 $\pm$   0.06 & -0.34 &   3 &  -6.13 $\pm$   0.01 &   2 \\                       
&  LTE &   -6.17 $\pm$   0.10 &   5 &  -6.28 $\pm$   0.07 & -0.57 &   3 &  -6.25 $\pm$   0.04 &   2 \\                       
HD~72660 & NLTE &   -5.38 $\pm$   0.06 &  14 &  -5.45 $\pm$   0.10 &  0.26 &   9 &  -5.53 $\pm$   0.12 &   3 \\                       
&  LTE &   -5.45 $\pm$   0.07 &  14 &  -5.50 $\pm$   0.11 &  0.21 &   9 &  -5.33 $\pm$   0.18 &   3 \\                       
HD~145788& NLTE &   -5.23 $\pm$   0.03 &   5 &  -5.33 $\pm$   0.07 &  0.38 &   8 &                     &     \\                       
&  LTE &   -5.56 $\pm$   0.08 &   5 &  -5.64 $\pm$   0.12 &  0.07 &   8 &                     &     \\                       
Siruis   & NLTE &   -5.93 $\pm$   0.03 &   5 &  -5.92 $\pm$   0.03 & -0.21 &   4 &  -6.04 $\pm$   0.04 &   3 \\                       
&  LTE &   -6.14 $\pm$   0.08 &   5 &  -6.12 $\pm$   0.06 & -0.41 &   4 &  -6.24 $\pm$   0.08 &   3 \\                       
21~Peg   & NLTE &   -5.49  &   1 &  -5.59 $\pm$   0.08 &  0.12 &   4 &  -5.81 $\pm$   0.04 &   2 \\                       
&  LTE &   -5.99  &   1 &  -6.00 $\pm$   0.09 & -0.29 &   4 &  -6.31 $\pm$   0.01 &   2 \\                       
HD~17081 & NLTE &                      &     &  -5.64  &  0.07 &   1 &  -5.64 $\pm$   0.00 &   2 \\                       
&  LTE &                      &     &  -5.83  & -0.12 &   1 &  -6.44 $\pm$   0.00 &   2 \\                       
HD~160762& NLTE &                      &     &  -5.45  &  0.26 &   1 &  -5.37 $\pm$   0.13 &   3 \\                       
&  LTE &                      &     &  -5.26  &  0.45 &   1 &   \multicolumn{2}{l}{emission} \\                  
			\hline
\multicolumn{9}{l}{$\sigma = \sqrt{ \Sigma (x - x_i )^2 /(N - 1)}$}\\
\multicolumn{9}{l}{ solar calcium abundance log A$_{\odot}$ = --5.71 \citep{Lodders2009}	}	\\
			\hline
		\end{tabular}
\end{table*}

\subsection{Calcium NLTE abundance determination with SME code}

As it was written in Section~\ref{se}, we calculated departure coefficients of Ca\ione\ and Ca\ii\ levels for a grid of LLmodels atmospheres \citep{llmod}. Earlier, the same calculations were performed for grids of MARCS plane-parallel models \citep{marcs}. These NLTE data are implemented in the current version of the SME (Spectroscopy Made Easy) package of automatic parameters and abundance determination based on high-resolution stellar spectra \citep{2017A&A...597A..16P}. Atmospheric parameters and element abundances in SME are derived by the interpolation inside the grid models. For cool stars (MARCS model grids) the accuracy of authomatic NLTE Ca abundance determinations was tested in the paper by \citet[][see Figs.~1 and 3 of this paper]{2017ASPC..510..509P}. Here, we tested the automatic method for hotter stars with HD~32115 and HD~72660. 
For HD~32115 the SME abundances are -5.68$\pm$0.07 (Ca\ione) and -5.60$\pm$0.14 (Ca\ii). For HD~72660 the corresponding values are -5.44$\pm$0.13 (Ca\ione) and -5.52$\pm$0.10 (Ca\ii), thus fully consists with the line-by-line NLTE abundance determinations (see Table~\ref{abund}). The larger error bars derived with SME are due to additional uncertainties caused by spectral noise which is included in the SME procedure. 
 Our results allow to increase the range of stellar atmospheric parameters where we can carry out the NLTE calcium abundance determinations using automatic methods.

\section{Conclusions}
\label{con}

In this study, the NLTE line formation for Ca\ione--\ii\ is investigated for the first time in the stellar parameter range corresponding to BAF-type stars. 
 We use a fairly complete model atom of Ca\ione--\ii\ from \citet{mash_ca}.
The NLTE abundance corrections for lines of   Ca\ione\ and Ca\ii\  can be either positive or negative, depending on the line and atmospheric parameters.

Using nine stars, with well-determined atmospheric parameters, which cover the \teff\ range from 7250 K to 17500 K, and high resolution and high S/N spectra available, we prove that NLTE leads to consistent abundances from different lines of the two ionisation stages, Ca\ione\ and Ca\ii.
To be precise, the NLTE abundances from Ca\ii\ lines in the visible and IR (8912-27, 9890 \AA) spectral range agree within 0.1 dex for each star, in contrast to LTE, where the abundance discrepancy between these two groups of lines ranges from   $-0.5$~dex to 0.6 dex  for different stars.
In case of 21 Peg, NLTE reduces the abundance difference compared to LTE, however, it remains still large and amounts to 0.2 dex. 
We obtain that, for each star, the average abundance difference between Ca\ione\ and Ca\ii\ does not exceed 0.1 dex, independent of either NLTE or LTE. 
However, the NLTE abundances are higher and NLTE always leads to substantially smaller line-to-line abundance scatter compared to LTE.

We do not recommend to use the Ca\ii\ IR triplet lines (8498, 8542, 8662 \AA) for accurate abundance determination due to difficulties with continuum placement of the observed spectrum and uncertainty of background opacity calculations near the Paschen jump.
It is worth noting that the Gaia spectrograph focuses exactly on that spectral region and the Ca\ii\ IR triplet lines are going to be {\bf employed} to determine calcium abundances of AB type stars. We caution against using Ca\ii\ 8498 and 8542~\AA\ and recommend to use 8662~\AA, given that an observed spectrum normalisation is done properly.    

Our NLTE method reproduces the  Ca\ii\ 8912-27, 9890 \AA\ lines observed in  emission in the late B-type star HD~160762  with the classical plane-parallel and LTE model atmosphere.

We provide NLTE abundance corrections for lines of Ca\ione\ and Ca\ii\ in the following range of atmospheric parameters: 7000~K $\leq$ \teff\ $\leq$ 13000~K, 3.2 $\leq$ log~g $\leq$ 5.0, $-0.5 \leq$ [Fe/H] $\leq 0.5$, \vt = 2.0 \kms.

Our theoretical NLTE results can be useful for studies of calcium abundance of Am stars. 
For example, a correlation found in LTE between calcium abundance and effective temperature,
in the sense that the cooler objects are more Ca-deficient than the hotter ones \citep{1998A&A...330..651K}, is expected to be strengthened in NLTE. This can be useful to constrain the models of diffusion in Am stars from observations.

\section*{Acknowledgments}

We thank V. Khalack for providing spectrum of HD~72660 that was obtained with ESPaDOnS at the CFHT and
A. Korn for providing spectrum of Vega that was obtained with FOCES at the Calar Alto 2.2m telescope.
This research is based on observations obtained with MegaPrime/MegaCam, a joint project of CFHT and CEA/IRFU, at the Canada--France--Hawaii
Telescope (CFHT) which is operated by the National Research Council (NRC) of Canada, the Institut National des
Science de l'Univers of the Centre National de la Recherche Scientifique (CNRS) of France, and the University of Hawaii.
The operations at the Canada-France-Hawaii Telescope are conducted with care and respect from the summit of Maunakea which is a significant cultural and historic site.
We used the DETAIL code provided by K. Butler, a participant of the "Non-LTE Line Formation for Trace Elements in Stellar Atmospheres" School
(July 29-August 2, 2007, Nice, France).
TS and LM are grateful to the Russian Scientific Foundation (grant 17-13-01144 supported via Herzen University), and
TR is grateful to the Russian Foundation for Basic Research (grant 15-02-06046).
We also thank the referee, Luca Fossati, for useful comments and suggestions.
We made use of the NIST, SIMBAD, and VALD databases.

\bibliography{ti,titanium,atomic_data,nlte1,mashonkina,stellar_parameter1,stellar_parameter,stellar_parameter2}
\bibliographystyle{mn2e}

\bsp

\label{lastpage}

\end{document}